%% file: main.tex
\documentclass[aps, prx, reprint, superscriptaddress]{revtex4-2}

\usepackage{amsmath}
\usepackage{amssymb}
\usepackage{graphicx}
\usepackage[colorlinks=true, allcolors=blue]{hyperref}

\DeclareUnicodeCharacter{2215}{/}

\begin{document}

    \title{Spin Grid States for Quantum Metrology in Atomic Clocks Limited by Spontaneous Emission}

\author{Marius Burgath}
\email{marius.burgath@itp.uni-hannover.de} 
\affiliation{Institute for Theoretical Physics, Leibniz Universität Hannover, Appelstraße 2, 30167 Hannover, Germany}
\affiliation{Institute for Theoretical Physics, University of Innsbruck, Technikerstr. 21a, 6020 Innsbruck, Austria} 

\author{Klemens Hammerer}
\email{klemens.hammerer@uibk.ac.at}
\affiliation{Institute for Theoretical Physics, University of Innsbruck, Technikerstr. 21a, 6020 Innsbruck, Austria}
\affiliation{Institute for Quantum Optics and Quantum Information of the Austrian Academy of Sciences, 6020 Innsbruck, Austria} 
\affiliation{Institute for Theoretical Physics, Leibniz Universität Hannover, Appelstraße 2, 30167 Hannover, Germany}

\date{\today}

\begin{abstract}
Entanglement can enhance precision in phase estimation and in frequency estimation with atomic clocks, but it remains a central challenge to identify useful states and measurements under realistic noise processes. Here, we study quantum metrology with an ensemble of atoms subject to spontaneous emission, which limits frequency estimation by constraining the useful interrogation time. We identify spin grid states (SGSs) as the relevant near-optimal probes beyond a crossover at an ensemble size of $51$, where GHZ-like states cease to be optimal. SGSs display a periodic grid on the Bloch sphere and can be generated with two one-axis-twisting (OAT) operations separated by a collective rotation. By optimizing the quantum Fisher information over permutationally symmetric states, we find that SGSs perform close to the globally optimal probes for intermediate and large ensembles, which turn out to be spin GKP states. We further present a sequential readout strategy that nearly saturates the corresponding quantum Fisher information. This strategy uses an OAT echo to map spontaneous-emission events onto collective rotations, making the corresponding jump sectors distinguishable and allowing the measurement basis to be conditioned on the number of decay events. Together, these results show that SGSs provide a practical route to quantum-enhanced phase and frequency metrology in atomic ensembles limited by spontaneous emission.
\end{abstract}

\maketitle 

\input{MAIN/Introduction}

\input{MAIN/DEF_SGS}

\input{MAIN/SGS_PhaseEst}

\input{MAIN/FreqMet_SGS}

\input{MAIN/Saturating_Bound}

\input{MAIN/Conclusion_Outlook}

\bibliography{bibliography}

\begin{widetext}

\section*{Methods and Supplementary Material}

\input{SUPP_MAT/1_OAT_as_rotations}

\input{SUPP_MAT/2_Model_Basics}

\input{SUPP_MAT/3_PermInv}

\input{SUPP_MAT/4_GlobalOpt}

\input{SUPP_MAT/5_GKP_SpinGKP}

\input{SUPP_MAT/6_SGS_prot}

\input{SUPP_MAT/7_StateDist}

\end{widetext}

\end{document}

%% file: MAIN/Introduction.tex
\section{Introduction}
Quantum superposition and entanglement enable measurement precision beyond classical limits in platforms ranging from atomic clocks \cite{Kaufm25, Dietze26, Pedrozo20, Robinson24} to sensors for fundamental physics \cite{Kimball23, DarkMatter_CAT} and gravitational-wave detectors \cite{Goda08}. For uncorrelated atoms or photons, quantum projection noise sets the standard quantum limit (SQL), which can be surpassed with entangled probes such as spin-squeezed states (SSSs) \cite{Kitag93,Schulte20TWO, Pezze18, Robinson24} or GHZ states \cite{Bollinger96, Kiel24, Dietze26}. Yet the same quantum correlations that improve sensitivity are often fragile under noise and decoherence, making ultimate bounds such as the Heisenberg limit unattainable in realistic settings \cite{Demkowi12, Escher11, Knysh24, Jarzyna15}.

Motivated by current efforts to build next-generation optical atomic clocks \cite{Kaufm25, Dietze26}, we study metrology with an ensemble of atoms subject to spontaneous decay. Apart from quantum projection noise itself, spontaneous emission is arguably the most fundamental noise contribution in such systems. An atomic clock is based on probing the atoms with a laser for some time, and relies on extending this interrogation time to improve precision and frequency stability. At some point, however, the finite lifetime of the excited state imposes a restriction on the achievable precision. For certain clock ions, we are already entering this lifetime-limited regime, where the clock lasers are so coherent that their coherence time exceeds the lifetime of the probed atoms, for example for $\text{Ca}^{+}$ ions with $\tau_{\text{life}} = 1.1 \, \text{s}$ \cite{ClocksGuide, Kiel24}. In this regime, GHZ states provide a substantially improved measurement precision over classical strategies for small ensembles \cite{Kiel24}. This is surprising, since GHZ states cannot provide any quantum advantage over uncorrelated states for local dephasing noise \cite{Huelga97, Froewis14}. For larger atomic ensembles, however, the ultimate bounds of metrology limited by spontaneous emission, and the entangled states and tailored measurement strategies needed to approach those bounds, remain open. In this work, we address these questions and demonstrate how to harness entanglement for spontaneous-decay-limited metrology.

To identify the best achievable precision under spontaneous emission, we numerically maximize the quantum Fisher information over permutationally symmetric states for ensembles of up to $125$ atoms. Beyond the small-ensemble regime where GHZ-like states are optimal, this optimization yields spin GKP states \cite{Omanakuttan23} as the optimal probes. Since such states generally require comparatively complex preparation protocols \cite{Gutman24}, we ask whether the same metrological advantage can be obtained with much simpler resources native to atomic-clock platforms. We show that two one-axis-twisting (OAT) operations \cite{Kitag93} separated by a collective rotation generate spin grid states, whose Wigner functions display a periodic grid structure similar to that of spin GKP states. These states perform close to the symmetric-state optimum and provide a gain over SSSs. We further show that their large quantum Fisher information can be accessed for phase estimation and frequency metrology in atomic clocks by a sequential readout strategy. This strategy uses an OAT echo to map spontaneous-emission events onto collective rotations, making the corresponding jump sectors distinguishable and allowing the measurement basis to be conditioned on the number of decay events.

%% file: MAIN/DEF_SGS.tex
\section{Spin Grid States}
\begin{figure*}[t]
    \centering
    \includegraphics[width=2\columnwidth]{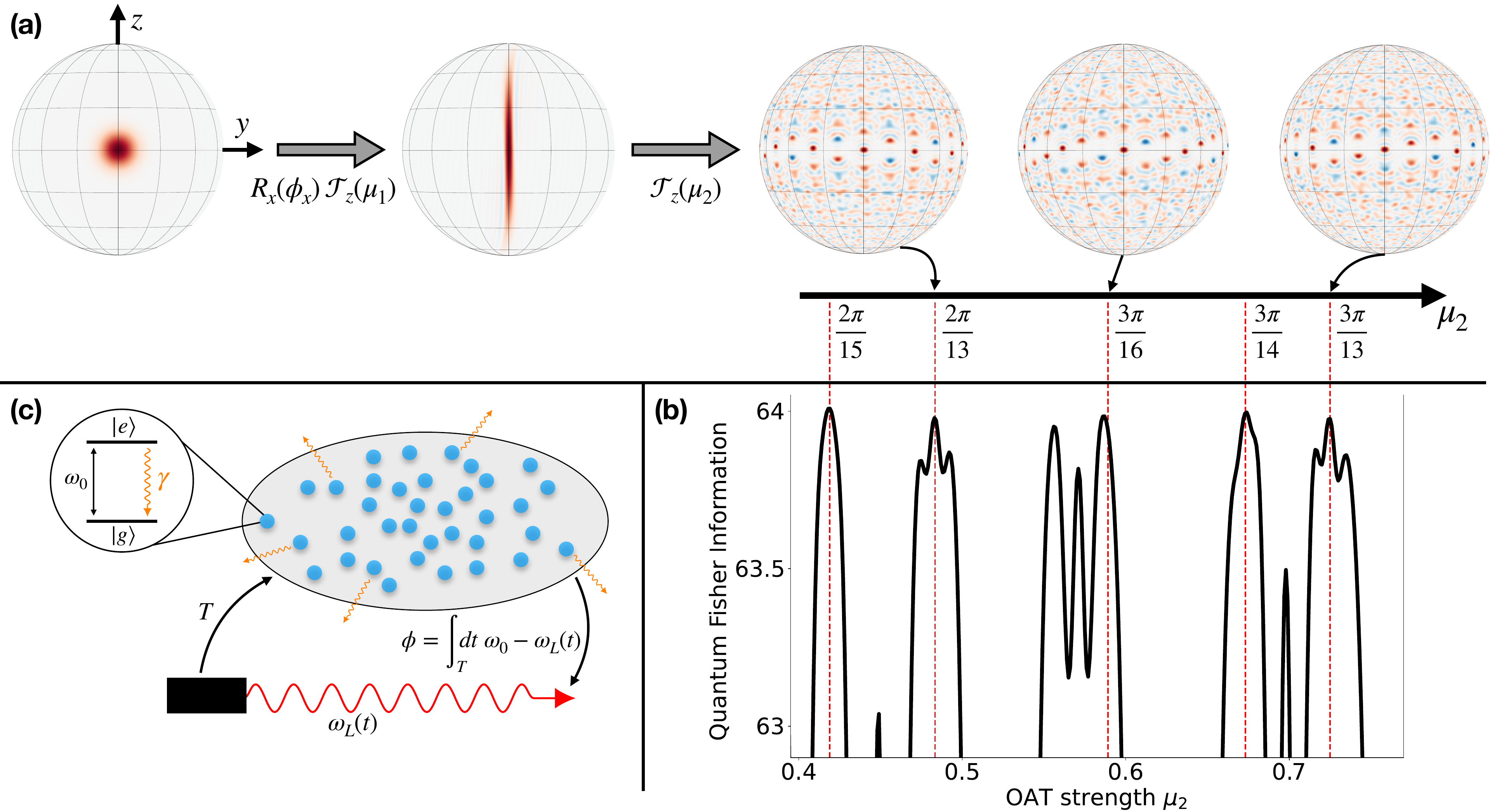}
    \caption{\textbf{(a)} Creation of SGSs in two steps, shown for $N=90$. First, an SSS is generated from a CSS by OAT and aligned by a collective rotation; the example shown uses $\mu_1 = 0.054$ and $\phi_x = -1.35$. Second, applying another OAT operation with strength $\mu_2$ superposes rotated copies of the SSS and produces grid structures in the spin Wigner function on the Bloch sphere. For rational values of $\mu_2$, the number and phases of these copies are determined by Eq. (\ref{eq:decomposeOATROT}). \textbf{(b)} Quantum Fisher information for phase estimation after spontaneous emission with $\eta = 0.25$, using the same fixed $\mu_1$ and $\phi_x$ as in (a). Local maxima occur near the rational twisting angles where SGSs form. \textbf{(c)} Atomic-clock setting: an ensemble of $N$ atoms undergoes spontaneous decay at rate $\gamma$ during an interrogation time $T$. The detuning between the laser frequency $\omega_L(t)$ and the atomic resonance frequency $\omega_0$ imprints the phase $\phi = \int_0^T [\omega_L(t)-\omega_0]\,dt$, which is estimated by a final measurement and used to stabilize the laser.}
    \label{Fig:1}
\end{figure*}
We consider an ensemble of $N$ identical atoms modeled as two-level systems (qubits), see Fig. \ref{Fig:1} (c). A natural resource for creating entanglement in such systems is OAT, $\mathcal{T}_{z}(\mu) = \exp \left( i \mu S_z^2 \right)$, where $S_z$ is the collective spin projection along $z$. A paradigmatic and metrologically important class of states generated by OAT is given by spin-squeezed states (SSSs) \cite{Kitag93}, which can be obtained by applying $\mathcal{T}_{z}(\mu)$ with $\mu \ll 1$ to a coherent spin state (CSS). Here, the CSS is defined as the $x$-polarized state $\lvert \text{CSS} \rangle = \left((\lvert g \rangle + \lvert e \rangle)/\sqrt{2}\right)^{\otimes N}$. The OAT operator describes an all-to-all pairwise interaction and can be routinely engineered in various experimental setups \cite{Pezze18, Bohnet16, Gross10, Strobel14, Eckner23}. For example, M\o{}lmer-S\o{}rensen gates \cite{Mölmer00} have been used to realize collective OAT and robustly create GHZ states of up to 24 calcium ions, corresponding to a gate with $\mu = \pi/2$ \cite{Pogorelov21}. Trapped beryllium ions have further been used to generate OAT and produce SSSs of more than $200$ ions as well as non-Gaussian, oversqueezed states \cite{Bohnet16}.

An efficient route towards non-Gaussian states is provided by the periodicity of OAT at rational twisting angles. For $\mu = p\pi/l$ with coprime $p, l \in \mathbb{N}$, the OAT operator can be decomposed into a finite sum of collective rotations $\mathcal{R}_z(\theta) = e^{-i \theta S_z}$,
\begin{equation}
    \label{eq:decomposeOATROT}
    \mathcal{T}_{z}(p \, \pi/l) = \sum_{m \in \mathcal{S}_{l,p}} \frac{1}{\sqrt{l}} \; e^{i \chi_{\tiny m}} \; \mathcal{R}_z\left(m \,\pi/ l\right).
\end{equation}
We derive this operator identity in Methods \hyperref[Meth:M1]{M1}; related decompositions have appeared previously for particular states in the context of fractional revivals \cite{Dooley14, Averbukh89, Agarwal93, Agarwal97}. Here, $\mathcal{S}_{l,p}$ denotes the even or odd integers in $\{0,1,\dots,2l-1\}$, respectively, depending on whether $p(l-N)$ is even or odd. The relative phases $\chi_{\tiny m}$ can be obtained by evaluating quadratic Gauss sums \cite{Dooley14, Woelk11} and generally depend on $p$, $l$, and $N$. Since the rotation angles span a whole period of $2 \pi$, applying $\mathcal{T}_{z}(p \, \pi/l)$ to any state generates a coherent superposition of $l$ copies of that state rotated about the $z$ axis. Notably, when the input state is an SSS, this superposition creates a periodic grid pattern in phase space for appropriate $l$, see Fig. \ref{Fig:1} (a). This grid structure extends fully around the equator of the Bloch sphere. The best grids are obtained if the SSS is weakly squeezed in $y$-direction with a twisting angle $\mu_1 \approx 0.9 \; \mu_{\text{min}}$, where $\mu_{\text{min}}$ minimizes the width $\Delta S_y (\mu)$ of the squeezed axis, see Methods \hyperref[Meth:M1]{M1}. Furthermore, $l \approx \pi/\Delta S_y(\mu_1)$ is a suitable choice to make sure the superposed states generate a grid. This feature is not fine tuned, but persists over a finite region of twisting angles around these values. Accordingly, we refer to states of the form $\lvert \psi_0 \rangle = \mathcal{T}_{z}(\mu_2) \, \mathcal{R}_x(\phi_x) \, \mathcal{T}_{z}(\mu_1) \lvert \text{CSS} \rangle$ with the aforementioned gate parameters as spin grid states (SGSs), where $\mathcal{R}_x(\phi_x) = e^{-i \phi_x S_x}$ aligns the SSS before the second OAT operation.

The periodic grid of an SGS resembles the phase-space grid of Gottesman-Kitaev-Preskill (GKP) states known from bosonic error correction \cite{Gottesman01, Flüh19, Takase23}. Bosonic GKP states are superpositions of displaced quadrature-squeezed states, while SGSs are superpositions of rotated SSSs. By analogy to bosonic GKP states, spin GKP states have been defined as polarized spin states with a grid structure on the Bloch sphere \cite{Omanakuttan23}. Such states can be prepared using sequences of squeezing operations and collective rotations, which are universal for permutationally symmetric pure states with a number of gates that scales polynomially in $N$ \cite{Bond25, Gutman24}. In particular, Ref. \cite{Gutman24} constructs spin GKP states using generalized two-axis-twisting operations, requiring more than ten such gates already for systems of $30$ qubits. SGSs are distinct from spin GKP states in that they have vanishing mean polarization, and their practical advantage is that a related grid structure can be generated with only two OAT operations and one collective rotation.

Figure \ref{Fig:1} (b) gives a first illustration of why these grid states are useful for noisy metrology under spontaneous emission. We plot the quantum Fisher information for phase estimation, a measure of the ultimate sensitivity of the final state to small changes of the imprinted phase $\phi$, while varying the second OAT strength $\mu_2$. The peaks of this sensitivity occur precisely near the rational values of $\mu_2$ for which the Wigner function forms a grid. Thus, the same periodic structure that follows from the OAT decomposition also marks the metrologically useful states. The following sections make this connection quantitative by comparing SGSs to the best states found by numerical optimization, first for the general problem of phase estimation and then for frequency estimation in atomic clocks.

%% file: MAIN/SGS_PhaseEst.tex
\section{Spin Grid States for Phase Estimation}
\begin{figure*}[t]
    \centering
    \includegraphics[width=2\columnwidth]{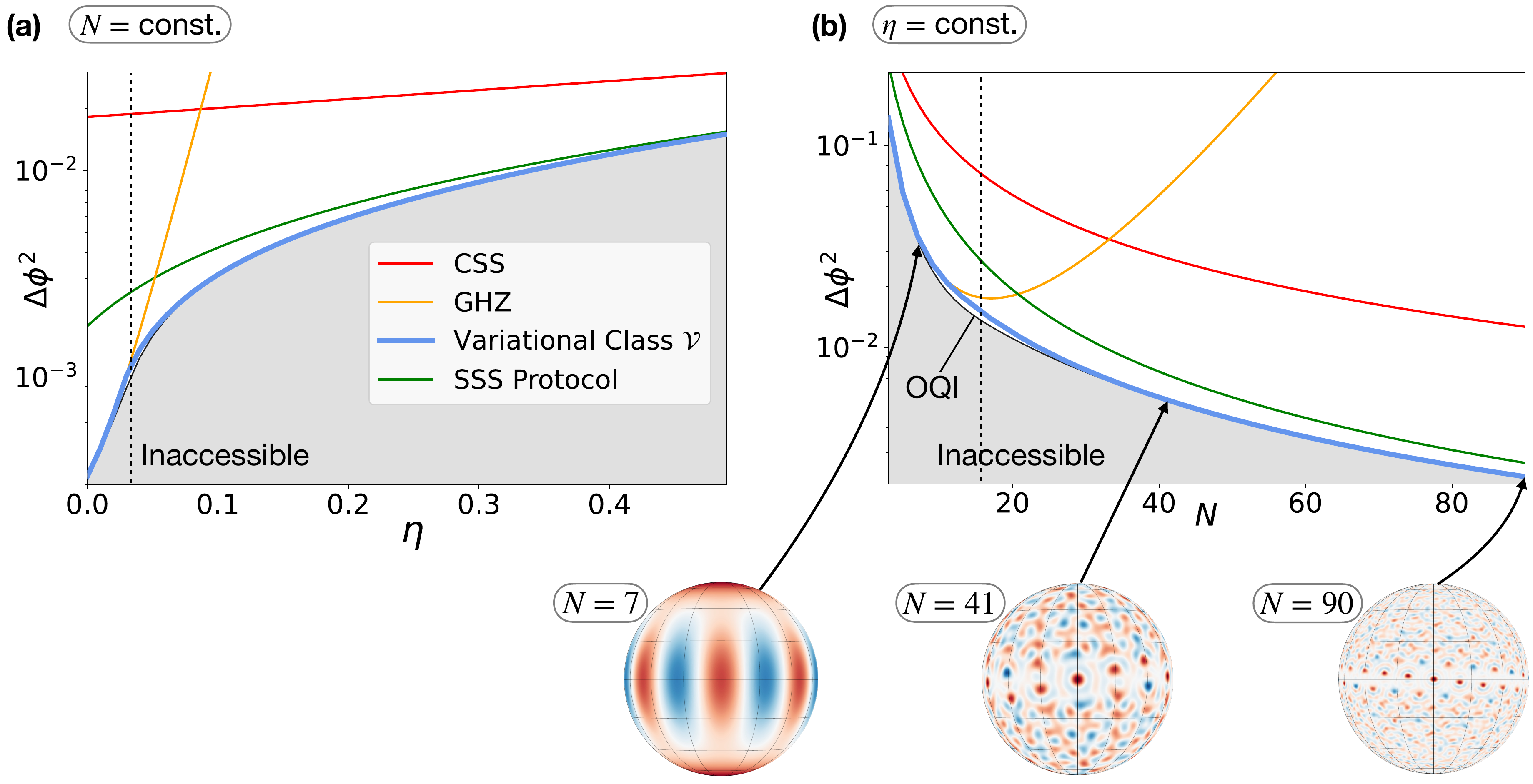}
    \caption{Phase-estimation variances $\Delta\phi^2$ for different initial states, shown on logarithmic axes. The CSS curve (red) gives the classical bound for uncorrelated atoms, while the OQI marks the optimum over permutationally symmetric initial states and bounds the inaccessible gray region. The variational optimum within $\mathcal{V}$ (blue) is compared to GHZ states (orange) and the optimized SSS protocol (green). \textbf{(a)} Fixed ensemble size $N=55$, with varying spontaneous-emission strength $\eta$. The dotted line indicates the crossover within $\mathcal{V}$ from GHZ-like states at weak noise to SGSs, and eventually SSS-like states, at stronger noise. \textbf{(b)} Fixed noise strength $\eta=0.13$, with varying ensemble size $N$. The optimum within $\mathcal{V}$ changes from GHZ-like states at small $N$ to SGSs for larger $N$, which remain close to the OQI and outperform the SSS protocol. For even $N$, GHZ-like states in $\mathcal{V}$ are obtained by replacing $\mathcal{R}_x$ with $\mathcal{R}_y$. The insets show Wigner functions of the corresponding optimized states on the Bloch sphere; the SGS grid becomes finer with increasing $N$.}
    \label{Fig:2}
\end{figure*}
Let $\rho_0$ denote the initial state of the atomic ensemble. Unitary evolution under a Hamiltonian $H \propto S_z$ rotates the state on the Bloch sphere and imprints a phase, $\rho_{\phi} = \mathcal{E}_{\phi}(\rho_0) = e^{-i \phi S_z} \rho_0 \, e^{i \phi S_z}$. Spontaneous emission is described by the noise channel $\mathcal{N}_{\eta}(\rho) = e^{\eta \mathcal{D}[\sigma_-]}(\rho)$, where $\mathcal{D}[\sigma_-](\rho) = \sum_i \sigma_-^{(i)} \rho \, \sigma_+^{(i)} - \left(\sigma_+^{(i)}\sigma_-^{(i)}\rho + \rho \, \sigma_+^{(i)}\sigma_-^{(i)}\right)/2$, $\sigma_- = \lvert g \rangle \! \langle e \rvert$, and $\eta$ denotes the dimensionless noise strength (Methods \hyperref[Meth:M2]{M2}). The noise channel commutes with the phase imprint, such that
\begin{equation}
    \label{eq:final_state}
    \rho_{\phi} (\eta) = \mathcal{E}_{\phi} \left( \mathcal{N}_{\eta}\left(\rho_0 \right) \right) = \mathcal{N}_{\eta} \left ( \mathcal{E}_{\phi} ( \rho_0) \right).    
\end{equation}
We then estimate the phase $\phi$ from the final state $\rho_{\phi}(\eta)$. According to the quantum Cramér-Rao bound (QCRB) \cite{Paris09, Kaufm25}, the variance of any locally unbiased estimator $\hat{\phi}$ is bounded by $\Delta \phi^2 \geq 1/\mathcal{F}_{\phi}$, where $\mathcal{F}_{\phi}$ is the quantum Fisher information (QFI), quantifying the sensitivity of $\rho_{\phi}(\eta)$ to small changes of $\phi$; see Methods \hyperref[Meth:M2]{M2}.

In Fig. \ref{Fig:2} we compare phase-estimation variances for several classes of initial states. As a benchmark, we define the optimal quantum interferometer (OQI) by numerically maximizing $\mathcal{F}_{\phi}$ over all permutationally symmetric initial states, which sets the boundary of the inaccessible region shown in gray. Exploiting the block-diagonal structure of $\rho_{\phi}(\eta)$, see Methods \hyperref[Meth:M3]{M3} and \hyperref[Meth:M4]{M4}, reduces the computational complexity to polynomial scaling in the system size and allows us to maximize $\mathcal{F}_{\phi}$ for $N \approx 100$ qubits on a standard laptop. To assess how close SGSs can come to this benchmark, we further optimize over the variational class
\begin{equation}
\begin{split}
    \label{eq:varclass}
    \mathcal{V} = \{ \, \mathcal{T}_{z}(\mu_2) \, \mathcal{R}_x(\phi_x) \, \mathcal{T}_{z}(\mu_1) \lvert \text{CSS} \rangle \, \},
\end{split}
\end{equation}
varying the gate parameters $(\mu_1, \phi_x, \mu_2)$. This class contains not only SGSs, but also CSSs, SSSs, and GHZ states for odd $N$; for even $N$, GHZ states can be obtained by replacing the $x$-rotation by a $y$-rotation. For all optimized states, we plot the QCRB variance $\Delta \phi^2 = 1/\mathcal{F}_{\phi}$. As a further reference, we compare to the ``SSS protocol'', namely the analytic estimator variance $\Delta \phi^2(\mu)$ of an SSS with squeezing along $y$ and a standard spin projective measurement \cite{Kiel24, Kitag93}, optimized over the squeezing strength $\mu$.

For a fixed ensemble size, Fig. \ref{Fig:2} (a) shows how the optimal state depends on the spontaneous-emission strength $\eta$. The variational class $\mathcal{V}$ closely follows the OQI over the full range of noise strengths, but different state classes are selected in different regimes. In the noiseless limit, GHZ states $\lvert \text{GHZ}\rangle = \left( \lvert g \rangle^{\otimes N} + \lvert e \rangle^{\otimes N} \right)/\sqrt{2}$ maximize the QFI and are therefore optimal \cite{Kaufm25}. Accordingly, for small $\eta$, the optimal states within $\mathcal{V}$ are GHZ-like. As the noise strength increases, however, GHZ states rapidly lose their advantage because their highly non-Gaussian multipartite coherence is fragile under decoherence. In the intermediate regime, approximately $0.1 \lesssim \eta \lesssim 0.4$ in Fig. \ref{Fig:2} (a), the optimal states within $\mathcal{V}$ are SGSs, which improve the precision over GHZ states by more than an order of magnitude and also outperform the SSS protocol. For still stronger noise, the grid structure is washed out by decoherence, and the optimum within $\mathcal{V}$ crosses over to SSS-like states, closing the gap to the SSS protocol.

Figure \ref{Fig:2} (b) shows the complementary case where the spontaneous-emission strength $\eta$ is fixed and the number of atoms is varied. The estimation uncertainty decreases with system size, while the optimal state again changes character. For small $N$, the OQI is well approximated by GHZ-like states. As $N$ increases, however, these states become more susceptible to spontaneous emission, and SGSs become optimal within $\mathcal{V}$. In this regime, the variational optimum remains close to the OQI and improves over the SSS protocol. For much larger $N$, the gap between the SSS protocol and the OQI is expected to close again, since SSSs are optimal in the asymptotic limit $N \rightarrow \infty$ \cite{Kiel24, Froewis14}. SGSs are the optimal states within $\mathcal{V}$ for all studied ensemble sizes beyond the small-$N$ GHZ-like regime, and remain close to the OQI throughout this range. They outperform the SSS protocol for all larger $N$; the advantage decreases asymptotically, but vanishes only in the limit $N \rightarrow \infty$.

%% file: MAIN/FreqMet_SGS.tex
\section{Frequency Estimation with Spin Grid States}
\begin{figure*}[t]
    \centering
    \includegraphics[width=2\columnwidth]{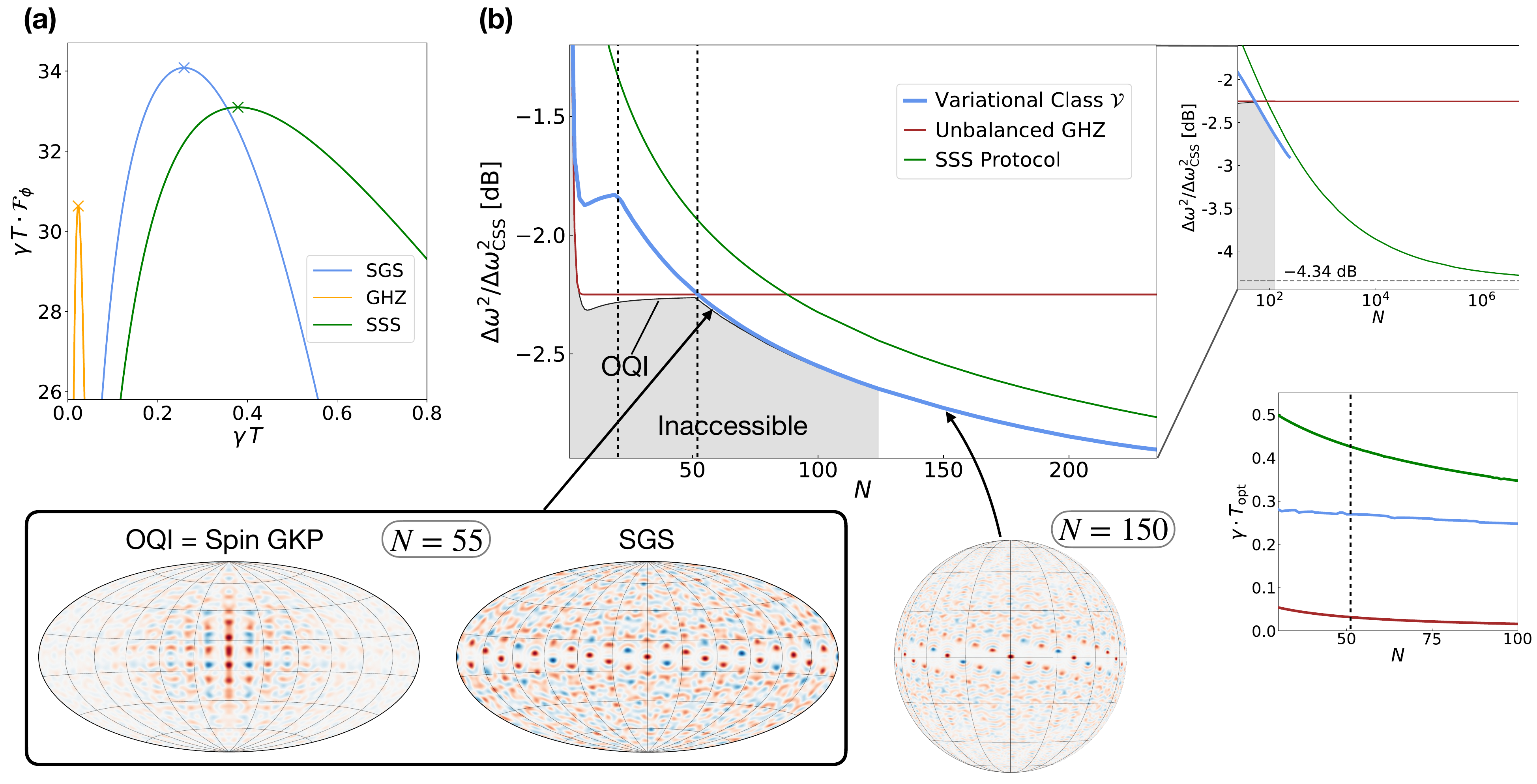}
    \caption{\textbf{(a)} Dependence of the optimal interrogation time on the initial state for $N=55$. The marked maxima represent the $\mathcal{F}_{\omega}$ values of the respective states. Here, the SGS is created with the gate parameters $(\mu_1, \phi_x, \mu_2) = (0.073, \, -0.939, \, 3 \pi/11)$. For the SSS, the squeezed axis is aligned along $y$ and the squeezing strength is $\mu = 0.049$. \textbf{(b)} Frequency estimation variances expressed as the gain in decibels over classical strategies. The OQI is defined as the ultimate lower bound across all permutationally symmetric initial states, so the grey shaded region is not accessible. Taking advantage of block diagonal structures, it was possible to numerically determine the OQI for ensembles with more than $100$ qubits. For $N=125$, it e.g. took less than $0.25$ seconds to evaluate the QFI for a given state on a standard laptop, making a numerical optimization feasible despite the large number of $N+1$ (complex) optimization parameters. We determined the OQI for up to $N = 125$, which is why the shaded area is not displayed for larger $N$. At $N=51$ (right dotted line), the OQI transitions between different optimal state classes. For $N \geq 51$, the initial states of the OQI are spin GKP states, which show a grid pattern in phase space. But unlike SGSs, these states are polarized, see the Hammer projections of the Bloch spheres in the boxed inset. For $N \lesssim 22$ (left dotted line), the optimal states within $\mathcal{V}$ are GHZ-like states, and for larger $N$ we obtained SGSs. \textbf{Upper right inset:} Optimality of SSSs for $N \rightarrow \infty$, with an asymptotic gain of $1/e$ over the SQL \cite{Knysh24}. \textbf{Lower right inset:} Each state has a particular optimal interrogation time, which is one order of magnitude larger for SGSs and SSSs than for unbalanced GHZ states. Hence, the optimal interrogation time of the OQI makes a jump at $N = 51$.}
    \label{Fig:3}
\end{figure*}
An atomic clock, as sketched in Fig. \ref{Fig:1} (c), is based on an ensemble of atoms that is repeatedly interrogated by a local oscillator (LO) laser in order to stabilize its frequency to the atomic transition $\omega_0$; the stabilized LO then provides the clock signal \cite{Kaufm25,Poli13,Schmidt15,Ye24,Mehlstaubler25, Marshall25}. Typically, this interrogation is performed using Ramsey interferometry \cite{Ramsey50}: Two laser pulses separated by a free evolution time $T$ imprint a phase $\phi = \int_0^T [\omega_L(t)-\omega_0]\,dt = \omega T$, where $\omega$ denotes the average detuning of the LO from the atomic transition during the Ramsey cycle. Let $\hat{\phi}$ be a phase estimator; then $\hat{\omega} = \hat{\phi}/T$ estimates this average detuning. The variance of $\hat{\omega}$ is $\Delta \phi^2 / T^2$, which suggests improving the clock's precision by increasing the duration $T$. In practice, however, spontaneous emission also accumulates during the interrogation and prevents the buildup of a coherent phase signal. Thus, the cycle duration is a control parameter to be optimized. Commonly, the stability of an atomic clock is quantified by the Allan deviation $\sigma_y(\tau)$ of the fractional frequency $y = \omega/\omega_0$ for a fixed averaging time $\tau = n T$, where $n$ is the number of independent cycles \cite{Kaufm25, Schmidt15, Kiel24}. For a clock limited by quantum projection noise, $\sigma_y(\tau)$ is related to the phase uncertainty by \cite{Kiel24} 
\begin{equation}
    \label{eq:Allan_dev}
    \sigma_y(\tau) = \frac{1}{\omega_0} \frac{\Delta \phi}{T \sqrt{n}} = \frac{1}{\omega_0} \frac{\Delta \phi}{\sqrt{T \tau}} =: \frac{\Delta \omega}{\omega_0}.
\end{equation}
The Allan deviation improves as $1/\sqrt{n}$ due to statistical averaging, giving a tradeoff between increased precision per interrogation and the number of shots within a fixed target averaging time $\tau$. Let $\tau_{\text{life}} = 1/{\gamma}$ be the atomic lifetime, such that $\eta = \gamma T$ is the spontaneous-emission strength and $\rho_{\phi}(\eta)$ the time-evolved state defined in Eq. (\ref{eq:final_state}). We use the QFI and set $\Delta \phi^{-2} = \mathcal{F}_{\phi}$ to identify ultimate limits of the Allan deviation. Since $\omega_0$, $\tau$, and $\gamma$ are fixed and identical irrespective of the initial state of the atomic ensemble, they can be treated as overall rescalings. Thus, $\sigma_y(\tau)^2 \propto \Delta \omega^2 \propto 1/(\eta \, \mathcal{F}_{\phi}(\eta))$, which has to be optimized over $\eta$. The relevant figure of merit for frequency estimation is therefore $\eta \, \mathcal{F}_{\phi}(\eta)$, optimized over the interrogation time, or equivalently over $\eta$. We denote this optimized quantity by $\mathcal{F}_{\omega} = \max_{\eta} \eta \, \mathcal{F}_{\phi}(\eta)$. Up to the overall rescaling by $\tau/\gamma$, its inverse gives the lower bound on the frequency variance.

Fig. \ref{Fig:3} (a) compares the optimized quantity $\eta \mathcal{F}_{\phi}(\eta)$ for different initial states and illustrates the corresponding optimal interrogation times. For $N=55$, the optimum of a GHZ state occurs at a much shorter time than that of an SGS. Evolving an SGS to its optimal time $T_{\text{opt}}$, we find that spontaneous emission shifts the grid slightly towards the south pole of the Bloch sphere, while the grid structure remains clearly visible. This persistence under local spontaneous emission explains why SGSs can be interrogated for moderate times, where GHZ states are already strongly decohered. Related self-similarity under local noise channels has also been observed and exploited for spin GKP states in quantum error correction \cite{oma26}.

In Fig. \ref{Fig:3} (b) we show the frequency-estimation bounds for different initial state classes. The OQI refers to a numerical maximization of $\mathcal{F}_{\omega}$ over all permutationally symmetric initial states, see Methods \hyperref[Meth:M3]{M3} and \hyperref[Meth:M4]{M4}. Furthermore, we maximized $\mathcal{F}_{\omega}$ within the variational class in Eq. (\ref{eq:varclass}), which was possible for up to $N \approx 235$ because $\mathcal{V}$ is parametrized by only the three gate parameters $(\mu_1,\phi_x,\mu_2)$. These results are expressed as lower bounds on the frequency variance, $\Delta \omega^2 = 1/\mathcal{F}_{\omega}$. We further compare to the analytic expression $\Delta \omega^2(\mu)$ of a conventional Ramsey protocol with SSSs, minimized over the squeezing parameter $\mu$ \cite{Kiel24, Kitag93}. Since asymptotically only a constant quantum gain over the SQL is possible \cite{Escher11, Knysh24, Kiel24}, we normalize all results by the metrological bound of classical strategies. This corresponds to the variance of a Ramsey protocol with CSSs, i.e. $\Delta \omega^2_{\text{CSS}} = e/N$.

We observe a transition between different optimal state classes at $N = 51$. For $N \geq 51$, the OQI is approximated very well by $\mathcal{V}$, and SGSs are the optimal states within the variational class. The initial states of the OQI also show a grid structure and can be identified as spin GKP states \cite{Omanakuttan23}, see the boxed inset in Fig. \ref{Fig:3} (b). A detailed discussion of spin GKP states and their optimality is given in Methods \hyperref[Meth:M5]{M5}. Globally, the Wigner functions of spin GKP states and SGSs are very different: SGSs extend around the full equator of the Bloch sphere, whereas spin GKP states are polarized. In the local frequentist setting considered here, however, the relevant feature is the local lattice structure near the working point. The optimization selects states with separated grid peaks in this region, which yield an improved lower bound compared to SSSs, see the gap to the green SSS-protocol curve. This is very different from local dephasing noise, where no states provide a noticeable gain over SSSs for any system size \cite{Froewis14}. For much larger $N$, the gain over SSSs decreases asymptotically, since SSSs are optimal in the limit $N \rightarrow \infty$ \cite{Kiel24, Froewis14}, see the upper inset in Fig. \ref{Fig:3} (b).

For $N \lesssim 50$, the frequency optimum is governed by the unbalanced GHZ states identified in Ref. \cite{Kiel24}, see the brown line. These states close the gap to the OQI for small ensembles, yield a substantial gain over SSSs, and admit an analytic measurement that saturates the QCRB \cite{Kiel24}. Note that for multi-level atoms, the same stability gain can be reached without the use of entangled states \cite{Hume26}. Unbalanced GHZ states already reach their asymptotic scaling $\Delta \omega^2 \propto 1/N$ at small $N$, and thus are far from optimal for large systems. In contrast, the quantum gain of SGSs over classical strategies increases with system size, and SGSs become the optimal states within $\mathcal{V}$ beyond the small-$N$ regime. The transition of the OQI between these two optimal initial-state classes is visible in the lower right inset of Fig. \ref{Fig:3} (b), which shows the optimal interrogation times $T_{\text{opt}}$. SGSs have an optimal interrogation time that is about an order of magnitude longer than that of unbalanced GHZ states. The same change occurs for the OQI in the transition region, where $T_{\text{opt}}$ jumps at $N=51$, as indicated by the dashed line. While the physical value of $T_{\text{opt}}$ depends on the lifetime through $\eta_{\text{opt}}=\gamma T_{\text{opt}}$, the crossover at $N=51$ itself is independent of $\gamma$. It identifies the ensemble size at which grid-like states first become more advantageous than the unbalanced GHZ strategy within the symmetric subspace.

%% file: MAIN/Saturating_Bound.tex
\section{Saturating the Lower Bound}
The QFI involves an optimization over all possible measurements and therefore quantifies the metrological potential of a state, but it does not by itself provide an experimentally accessible readout. In many cases, it is sufficient to apply an entangling unitary $\mathcal{U}_{\text{ent}}$ before measuring the population difference \cite{Kaufm25, Kiel24, Pezze18, Pezze16}, such that the effective observable
\begin{equation}
    \label{eq:coll_obs}
    X =\mathcal{U}_{\text{ent}}^{\dagger} \, S_z \; \mathcal{U}_{\text{ent}}
\end{equation}
is measured. For SGSs, however, numerical maximizations of the classical Fisher information (FI) over observables of the form in Eq. (\ref{eq:coll_obs}), including deep circuits of rotations and OAT, do not approach the QCRB. This raises the question of how to saturate the metrological bounds of SGSs.

To answer this question, it is useful to first resolve the final state according to the number of spontaneous-emission events that occurred, see Methods \hyperref[Meth:M6]{M6}. The state in Eq. (\ref{eq:final_state}) can then be written as $\rho_{\phi}(\gamma T) = \sum_{n=0}^N \rho^{(n)}_{\phi}(\gamma T)$, where the unnormalized states conditioned on $n$ spontaneous-emission events are
\begin{equation}
\begin{split}
    \label{eq:unravel_basic}
    \rho^{(n)}_{\phi}(\gamma T)
    &=
    \frac{\left( 1 - e^{- \gamma T} \right)^n}{n!}
    e^{- i H_{\text{eff}} \, T}
    \mathcal{J}^n\left[ \sigma_{-} \right]\left( \rho_0 \right)
    e^{i H_{\text{eff}}^{\dagger}\,T}\\
    &=: \mathcal{E}^{(n)}(\rho_0)
\end{split}
\end{equation}
Here, $H_{\text{eff}} = (\omega - i\gamma/2)S_z - i\gamma N/4$ is the effective Hamiltonian, and the spontaneous-emission jumps are described by $\mathcal{J}\left[ \sigma_{-} \right]\left( \rho \right) = \sum_m \sigma_-^{(m)} \, \rho \, \sigma_+^{(m)}$. From now on, we assume that the initial state is an SGSs, i.e. $\lvert \psi_0 \rangle = \mathcal{T}_z(\mu_2) \lvert \xi \rangle$ with suitable $\mu_2$ and where $\lvert \xi \rangle=\mathcal{R}_x(\phi_x) \mathcal{T}_{z}(\mu_1) \lvert \text{CSS} \rangle$ is a SSS, cf. Fig.~\ref{Fig:1} (a). In this case, the unnormalized states $\rho^{(n)}_{\phi}(\gamma T)$ turn out to be almost perfectly pairwise orthogonal, as quantified by the state-distinguishability analysis in Methods \hyperref[Meth:M7]{M7}. There, we analyze the discrimination of the different $n$-jump sectors using the ``pretty good measurement'' (PGM) \cite{Hausladen94} and find that SGSs enable near-perfect distinguishability between states associated with different numbers of jumps. This is an intriguing feature, and generally not the case for CSSs, SSSs, or randomly generated initial states. It implies that phase estimation with SGSs benefits from knowing how many spontaneous-emission events occurred, so that a suitable measurement can be chosen in each sector.

Before showing how to gain information about the number $n$ of spontaneous-emission events, we first discuss suitable measurements under the assumption that the $n$-jump sector is known. If an OAT echo is applied at the end of the Ramsey sequence, the final state becomes $\mathcal{T}_{z}(\mu_2)^{\dagger} \,\rho_{\phi}(\gamma T) \, \mathcal{T}_{z}(\mu_2) = \sum_{n = 0}^N \bar{\rho}_{\phi}^{(n)}(\gamma T)$ with
\begin{equation}
\begin{split}
    \label{eq:njumpcontrib_NOECHO}
    \bar{\rho}_{\phi}^{(n)}(\gamma T)
    =
    e^{2i n \mu_2 S_z}
    \mathcal{E}^{(n)}\!\left(\lvert \xi \rangle \! \langle \xi \rvert\right)
    e^{-2i n \mu_2 S_z}.
\end{split}
\end{equation}
Remarkably, the effect of the OAT echo thus reduces, in each $n$-jump sector, to an additional collective rotation by the discrete angle $2n\mu_2$ about the $z$-axis. The state can therefore be viewed as a mixture of states obtained from the SSS $\lvert\xi\rangle$ by phase imprinting and exactly $n$ spontaneous-emission events, followed by this $n$-dependent rotation. For the gate parameters of SGSs, the rotation angle is moreover a rational multiple of $\pi$. The structure of Eq.~(\ref{eq:njumpcontrib_NOECHO}) follows from the commutation relation (Methods \hyperref[Meth:M6]{M6}) between OAT operations and spontaneous emission jumps,
\begin{equation}
\begin{split}
    &\mathcal{J} \left[ \sigma_{-} \right]\left( \mathcal{T}_{z}(\mu_2) \, \rho \, \mathcal{T}_{z}(\mu_2)^{\dagger}  \right) \\[5pt]
    & \quad \quad \quad =
    \mathcal{T}_{z}(\mu_2)
    \left ( \mathcal{J} \left[ e^{2 i \mu_2 S_z} \, \sigma_{-} \right](\rho) \right)
    \mathcal{T}_{z}(\mu_2)^{\dagger}.
\end{split}
\end{equation}
Consequently, once the sector $n$ is known, the remaining readout problem resembles phase estimation with a decohered SSS at an $n$-dependent working point on the Bloch sphere. A suitable measurement in each sector is therefore obtained by applying $n$-dependent collective rotations, possibly supported by an OAT with small twisting angle $\mu \ll 1$, and then measuring the population difference $S_z$.

\begin{figure}[t]
    \centering
    \includegraphics[width=1\columnwidth]{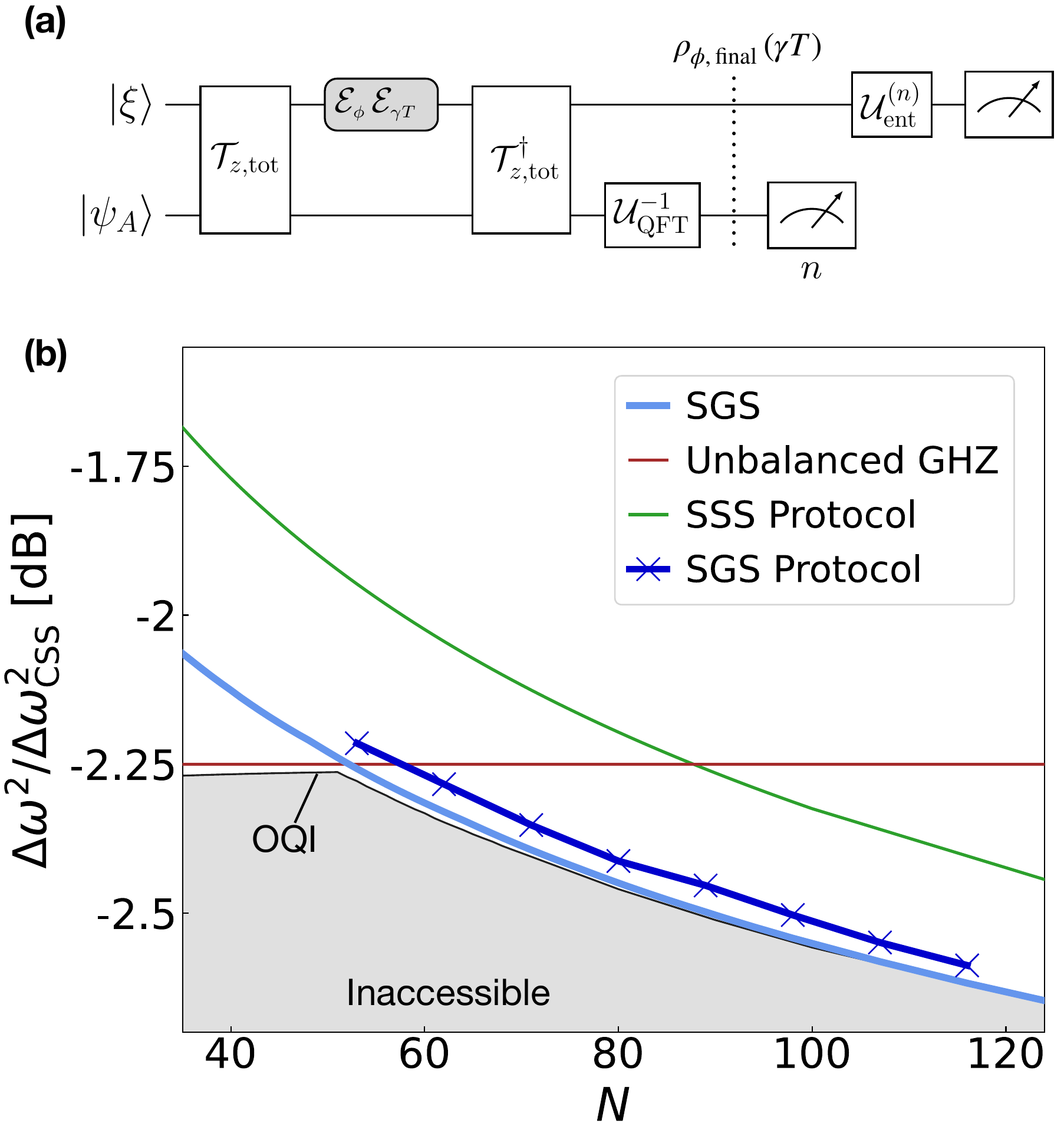}
    \caption{\textbf{(a)} Algorithm to estimate the phase with SGSs. Entanglement generated by $\mathcal{T}_{z, \text{tot}}$ encodes the number $n$ of decayed clock qubits onto the ancillary system. An inverse QFT is then used to map this information to a Dicke basis state. Measuring the spin projection $S_{z, A}$ determines $n$, and conditioned to this result an optimal entangling unitary $\mathcal{U}_{\text{ent}}^{(n)}$ is applied to estimate the phase. \textbf{(b)} Phase readout strategy compared to the frequency estimation bounds from Fig. \ref{Fig:3} (b). The ``SGS Protocol'' refers to the classical FI with respect to the algorithm in (a), optimized over the parameters of the gates $\mathcal{U}_{\text{ent}}^{(n)}$. We did the optimization for $N = 53, 62, 71, 80, 89, 98, 107,$ and $ 116$ clock qubits (marked by crosses). The parameters ($\mu_1, \phi_x, \mu_2$) were taken as the ones corresponding to the optimal SGSs in Fig. \ref{Fig:3} (b). We used ancillary systems with $n_0 \lesssim 0.4 \cdot \, N$. Including contributions $n > n_0$ led to neglegible improvements of the variance of the order $10^{-7}$. For $N = 53$, we e.g. took $n_0 = 20$.}
    \label{Fig:4}
\end{figure}

The remaining task is therefore to determine the $n$-jump sector without destroying the encoded phase information. We achieve this by engineering a QND measurement of $n$ with the help of an ancillary spin system. In the following, we refer to the system on which the phase is imprinted as the clock system. Fig. \ref{Fig:4} (a) shows the algorithm for the phase readout, which is discussed in detail in Methods \hyperref[Meth:M6]{M6}. We take an ancillary spin system with Dicke states $\lvert n \rangle$ such that $S_{z,A} \lvert n \rangle = (-j+n) \lvert n \rangle$ with $n = 0,1,2,\dots, 2j$. The combined system is prepared in the state $\lvert \xi \rangle \otimes \lvert \psi_A\rangle$, where $\lvert \psi_A \rangle$ is an equal superposition of the first $n_0$ Dicke states for some $n_0 \leq N$. Next, we apply the entangling operation $\mathcal{T}_{z,\, \text{tot}}(\mu_2) = \exp \left( {i \mu_2 (S_z + \epsilon S_{z,A})^2} \right)$ with a tunable parameter $\epsilon$. Subsequently, the clock atoms are subject to phase imprint and spontaneous emission. Due to the initial entanglement, the ancillary system receives phase kicks that depend on the number of spontaneous-emission events in the clock system. We then apply a $\mathcal{T}_{z,\, \text{tot}}(\mu_2)$ echo followed by an inverse quantum Fourier transformation (QFT) \cite{NielsenChuang} in the ancillary Dicke basis. For a suitable choice of $\epsilon$ and $n_0$, see Methods \hyperref[Meth:M6]{M6}, this inverse QFT maps the number of emission events onto an orthogonal ancilla state,
\begin{equation}
    \label{eq:final_ancillaTag}
    \rho_{\phi,\text{final}}(\gamma T) \approx \sum_{n=0}^{n_0-1}  \bar{\rho}^{(n)}_{\phi}(\gamma T) \, \otimes \, \lvert n \rangle \! \langle n \rvert.
\end{equation}
Measuring $S_{z,A}$ therefore determines the number of decay events while leaving the corresponding clock state $\bar{\rho}^{(n)}_{\phi}(\gamma T)$ intact. This realizes the desired QND measurement of the $n$-jump sector. Conditioned on the result $n$, we then apply the corresponding sector-dependent phase readout described above. Fig. \ref{Fig:4} (b) shows that this sequential strategy almost saturates the QFI of an SGS, and hence comes close to the global optimum across all measurements and initial states.

The ancillary system makes the $n$-jump sectors in Eq. (\ref{eq:final_ancillaTag}) perfectly distinguishable by construction. The fact that SGSs already have almost perfectly distinguishable jump contributions without ancillary degrees of freedom suggests that similar sequential strategies may also be possible on the clock system alone.

%% file: MAIN/Conclusion_Outlook.tex
\section{Conclusion and Outlook}
In the present work, we have analyzed phase and frequency estimation with an ensemble of atoms subject to spontaneous emission. We have found that, for intermediate to large ensembles, SGSs are metrologically near-optimal. These states show a grid structure on the Bloch sphere and can be created by a constant-depth circuit of two OAT operations and a collective rotation. Despite being entangled and highly non-Gaussian, SGSs are robust against spontaneous emission, which allows for Ramsey-interrogation times that are one order of magnitude larger than for GHZ states. This robustness stems from pairwise orthogonal $n$-jump sectors, which can be mapped to distinct positions on the Bloch sphere by an OAT echo without losing phase information. Once the number of spontaneous-emission events is known, the readout reduces to phase estimation with decohered SSSs.

Above an ensemble size of $N=51$, SGSs perform close to the OQI, whose states are in turn given by spin GKP states. Hence, a grid structure on the Bloch sphere can provide near-maximal QFI, and thus protects the stored phase information against spontaneous emission. From quantum computing in bosonic systems, it is known that quantum information can be protected against particle loss by using GKP qubits for the encoding \cite{Albert18, Harris25}. These codewords allow for a recovery map that restores logical states with high fidelity, because the particle-loss channel merely contracts GKP qubits towards the vacuum state without fundamentally destroying the grid structure even if many particles are lost. In a similar way, the Bloch-sphere grid of SGSs and spin GKP states is preserved even for large Ramsey-interrogation times, which protects the phase information stored in these states. However, the optimal recovery map used to restore bosonic GKP codes is complicated and practical decompositions into native operations available on a given experimental platform are not known. Likewise, for spin GKP states it is unclear how to access the phase information stored in these states. But for SGSs, we have identified a strategy for the phase readout that intuitively explains their robustness against spontaneous emission. The pairwise orthogonal $n$-jump sectors provide information about the number of decay events, such that an optimal measurement basis can be chosen for a particular number of spontaneous-emission events.

Spin GKP states have already been used for quantum error correction with collective spin systems \cite{Omanakuttan23}, which raises the question whether the grid pattern of SGSs can be exploited in a similar way. Furthermore, bosonic GKP states can be prepared by collective emission of spin GKP states \cite{Gutman24}. A natural question is whether a similar mapping of SGSs to the bosonic phase space still resembles a grid pattern. These connections suggest that SGSs may be useful more broadly as non-Gaussian resources in collective-spin and bosonic quantum systems, beyond their use for metrology.

\begin{acknowledgments}
We acknowledge discussions with P. O. Schmidt. We acknowledge support from DFG through the Collaborative Research Center SFB 1227 (DQ-mat, Project-ID 274200144).
\end{acknowledgments}

%% file: SUPP_MAT/1_OAT_as_rotations.tex
\section*{M1. \enskip Decomposition of the OAT operator}
\label{Meth:M1}
We take a spin system with angular momentum $j = N/2$ and Dicke eigenstates $\{\lvert n \rangle \}$ such that $S_z \lvert n \rangle = (N/2 - n) \lvert n \rangle$. Thus, the OAT operator $\mathcal{T}_z(\mu) = \exp(i \mu S_z^2)$ for some $\mu = p\pi/l$ with $p,l \in \mathbb{N}$ satisfies
\begin{equation}
    \mathcal{T}_z(p \pi/l) \lvert n \rangle = e^{i \frac{p \pi}{l} S_z^2} \lvert n \rangle = e^{i \frac{p \pi}{l} \left( \frac{N}{2} - n\right)^2} \lvert n \rangle.
\end{equation}
In the following, we will assume that the fraction is already simplified such that $p$ and $l$ are coprime numbers. The function $f_n := \exp \left(i \frac{p \pi}{l} \left( \frac{N}{2} - n\right)^2 \right)$ is periodic with $f_{n \pm 2l} = f_n$. Consequently, a discrete Fourier transformation \cite{Dooley14, Averbukh89, Agarwal93, Agarwal97} on a grid of length $2l$ yields
\begin{equation}
    e^{i \frac{p \pi}{l} \left( \frac{N}{2} - k \right)^2} = \frac{1}{2l} \sum_{m=0}^{2l-1} e^{i 2 \pi k \frac{m}{2l}}  \left( \sum_{r = 0}^{2l-1} e^{i \frac{p \pi}{l} \left( \frac{N}{2} - r \right)^2} \, e^{- i 2 \pi m  \frac{r}{2l}} \right) = \frac{1}{2l} \sum_{m=0}^{2l-1} e^{i 2 \pi k \frac{m}{2l}} \, Y_m,
\end{equation}
with Fourier expansion coefficients $Y_m$. Using rotations $\mathcal{R}_z(\theta) = \exp(-i \theta S_z)$, the OAT operator can be expressed as
\begin{equation}
    \mathcal{T}_z \left( p \pi/l \right) \lvert n \rangle = \left( \sum_{m = 0}^{2l-1} \frac{1}{2l} Y_m e^{i\frac{\pi m }{l} \frac{N}{2}} \mathcal{R}_z \left( \pi m/l \right)\right) \, \lvert n \rangle.
\end{equation}
Overall, the OAT operator decomposes into a sum of equally spaced rotations via $ \mathcal{T}_z \left( p \pi/l \right) = \sum_{m=0}^{2l-1} \; c_m \;  \mathcal{R}_z(\pi m/l)$. Particularly note that the angles of the $z$-rotations do not depend on the value $p$. For the coefficients $c_m$ we have
\begin{align}
    c_m &= \frac{1}{2l} Y_m e^{i \frac{\pi m}{l} \frac{N}{2}} = \frac{1}{2l} \sum_{r=0}^{2l-1} e^{i \frac{p \pi}{l} \left( \frac{N}{2} - r\right)^2} e^{- i \pi \frac{m r}{l}} e^{i \frac{\pi m}{l} \frac{N}{2}} = e^{i \frac{\pi m}{l} \frac{N}{2}} e^{i \frac{p \pi}{l} \frac{N^2}{4}} \, \left( \frac{1 + (-1)^{pl-K}}{2l} \right) \underbrace{\sum_{r=0}^{l-1} e^{i \frac{\pi}{l}\left( pr^2 - rK \right)} }_{=: A} \; \; \; \;,
\end{align}
where the abbreviation $K = m + Np$ has been defined. The sum $A$ has the form of a generalized quadratic Gauss sum \cite{Woelk11}. Note that the factor $1 + (-1)^{pl-K}$ is nonzero only if $pl-K$ even, which will be assumed in the following. Under this assumption, one can show that the norm of $A$ is given by $\lvert l \rvert$. We have:
\begin{align}
	\lvert A \rvert^2 &= \sum_{r=0}^{l-1} \;  \sum_{r^{\prime}=0}^{l-1} \; e^{i \frac{\pi}{l}\left( pr^2 - rK \right)} e^{i \frac{\pi}{l}\left( pr^{\prime 2} - r^{\prime}K \right)} = \sum_{r, r^{\prime} = 0}^{l-1} \; e^{i \frac{\pi p}{l}(r-r^{\prime})(r+r^{\prime})} \; e^{-i \frac{\pi}{l} K (r - r^{\prime})}.
\end{align}
Defining new summation variables by $u = r - r^{\prime}$ and $v = r + r^{\prime}$ one obtains
\begin{align}
	\lvert A \rvert^2 &= \sum_{u = -(l-1)}^{l-1} \;  \sum_{\substack{v = |u|\\\text{step size } 2 }}^{2(l-1)-|u|}\; e^{i \frac{p \pi}{l}u v} \; e^{-i \frac{\pi}{l} K u}  = l \; + \;  \text{Re} \left ( \sum_{u=1}^{l-1}  \;  \; \sum_{v = 0}^{l-1-u} \; e^{i \frac{p \pi}{l}u (2v+u)} \; e^{-i \frac{\pi}{l} K u} \right ) \\[7.5pt] &= \label{eq:transformed_summation} l \; + \;  \sum_{u=1}^{l-1} \; \;  \sum_{v = 0}^{l-1-u} \; \cos\left(\pi u (p(2v+u)-K)/l\right)  =: l \; + \; \sum_{u=1}^{l-1}  \mathcal{S}(u).
\end{align}
At the second equality, the first sum was split into sums over positive and negative $u$, and the value $l$ stems from the term with $u=0$. The second sum $\mathcal{S}(u)$ can be evaluated using the standard cosine progression formula, i.e.
\begin{equation}
	\sum_{k=0}^{n-1} \, \cos(a + kd) = \frac{\sin(nd/2)}{\sin(d/2)} \cos((2a + (n-1)d)/2).
\end{equation}
With $n = l-u$, $a = \pi u (pu-K)/l$, and $d = 2 \pi u p/l$ we find
\begin{equation}
	\mathcal{S}(u) = - \frac{\sin(u^2 p \pi/l)}{\sin(u p \pi/l)} \cos(\pi u (K + p)/l).
\end{equation}
Note the $p$ and $l$ are coprime natural numbers by assumption, such that $l$ does not divide $u p$ since $u < l$. So $u p /l$ is not an integer number, and thus $\sin(u p \pi/l) \neq 0$. For a shift $u \rightarrow l - u$ one can directly see that
\begin{equation}
	\mathcal{S}(l-u) = - \frac{(-1)^{lp}\sin(u^2 p \pi/l)}{(-1)^p (-1)\sin(u \pi p/l)} (-1)^{K+p} \cos(u \pi (K+p)/l) = - (-1)^{lp-K} \mathcal{S}(u) = - \mathcal{S}(u),
\end{equation}
since we look at the case where $pl-K = pl - (m + Np) = p(l-N) - m$ is an even number. So we have $\mathcal{S}(l-u) = -\mathcal{S}(u)$ such that the sum over $u$ in Eq. (\ref{eq:transformed_summation}) vanishes because the terms either cancel pairwise, or are zero. Thus, overall we have $\lvert c_m \rvert = 1/\sqrt{l}$ if $pl - m - Np$ is even, and $\lvert c_m \rvert = 0$ otherwise. Since $m \in \{0,1,\dots, 2l-1\}$, this means that every second $c_m$ vanishes, and $l$ coefficients are nontrivial. Consequently, the OAT operator is given by
\begin{equation}
    \mathcal{T}_{z} \left( p \pi/l \right) = \sum_{m \in S_{l,p}} \frac{1}{\sqrt{l}} \; e^{i \chi_{\tiny m}} \; \mathcal{R}_z(m \pi/l),
\end{equation}
Here, $\mathcal{S}_{l,p}$ denotes the even or odd integers in $\{0,1,\dots,2l-1\}$, respectively, depending on whether $p(l-N)$ is even or odd. So overall one can see that the OAT operation can be decomposed into a sum of $l$ rotations with equal weight and equal angular spacing, but different relative phases $\chi_{\tiny m}$. Note that the phases $\chi_{\tiny m}$ and their dependence on $p,l$, and $N$ can be determined using expressions for generalized quadratic Gauss sums \cite{Dooley14, Woelk11}.

A grid pattern on the Bloch sphere can be generated by applying $\mathcal{T}(p \pi/l)$ for suitable $l$ to an SSS given by $\lvert \xi \rangle = \mathcal{R}_x(\phi_x) \mathcal{T}_z(\mu_1) \lvert \text{CSS} \rangle$ with $\mu_1 \ll 1$, see Fig. \ref{Fig:1} (a). For a phase-space grid to form, a certain minimum size of the Bloch sphere is needed, which corresponds roughly to $N \gtrsim 30$. Furthermore, the squeezed axis of $\lvert \xi \rangle$ has to be approximately aligned in $y$-direction by choosing a suitable $\phi_x$, and the twisting angle should be slightly less than the angle $\mu_{\text{min}}(N)$ that minimizes the squeezing parameter $\zeta^2 := 4 \Delta S_y^2(\mu)/N$. Here, $\Delta S_y(\mu)$ denotes the uncertainty evaluated with respect to $\lvert \xi \rangle$. Maximizing the quantum Fisher information for $N \in [30, 230]$, we find that the optimal SGSs typically have $\mu_1 \approx 0.9 \; \mu_{\text{min}}(N)$ for all respective values of $N$. If $\lvert \xi \rangle$ were a slightly oversqueezed state instead, then the grid structure was already washed out and not visible. The Wigner negativities and the somewhat non-Gaussian shape of an oversqueezed state prevent the formation of a clearly visible grid structure. 

According to Eq. (\ref{eq:decomposeOATROT}), $l$ fixes the number of states that are superposed, whereas $p$ merely determines their relative phases. Interestingly, we obtained the best grid structures and the largest values of the quantum Fisher information by choosing $l \approx \pi/\Delta S_y(\mu_1)$. This choice of $l$ makes sure that the respective squeezed ellipses have no overlap on the Bloch sphere, but still provides enough constituents to interfere such that the grid structure can build up.

%% file: SUPP_MAT/2_Model_Basics.tex
\section*{M2. \enskip Noise Model and Quantum Estimation}
\label{Meth:M2}
Individual spontaneous emission is modelled by the master equation
\begin{equation}
    \label{eq:SE_MEQ}
    \dot{\rho} = \mathcal{D}[\sigma_-](\rho)  =  \frac{1}{2} \sum_{i=1}^N 2 \sigma_-^{(i)} \rho \, \sigma_+^{(i)} - \rho \, \sigma_+^{(i)} \sigma_-^{(i)} - \sigma_+^{(i)} \sigma_-^{(i)} \rho,
\end{equation}
where $\sigma_- = \lvert g \rangle \! \langle e \rvert$ is the Pauli step down operator of a single qubit. We further consider a collective rotation about the $z$-axis to imprint the phase $\phi$, which is a unitary dynamics generated by $\mathcal{L}_{z}(\rho) = - i \left [S_z, \rho \right ]$. Using the Pauli algebra $[\sigma_-, \sigma_z] = 2 \sigma_-$, one can easily show that both superoperators commute, i.e. $\left [ \mathcal{D}[\sigma_-], \mathcal{L}_z \right ] = 0$. So the final state $\rho_{\phi} (\eta)$ is given by Eq. (\ref{eq:final_state}), and can either be interpreted as imprinting the phase on a decohered mixed state, or applying the noise channel to a state which already contains phase information. For the particular case of frequency estimation, a phase $\phi = \omega T$ is imprinted during an interrogation cycle of duration $T$. Hence, if $\tau_{\text{life}} = 1/\gamma$ is the lifetime of the excited state, then the corresponding dimensionless noise strength is $\eta = \gamma T$.

A measurement $\{M_x \}$ on $\rho_{\phi} (\eta)$ gives result $x$ with probability $P(x \lvert\phi) = \text{tr} \left( M_x \,  \rho_{\phi} (\eta)\right)$. This result is used to obtain an estimate $\hat{\phi}(x)$ of the phase via an estimator function $\hat{\phi}$. We utilize the frequentist approach, so the phase is estimated locally around a working point $\phi_0$ \cite{Paris09}. For any locally unbiased estimator $\hat{\phi}$ and measurement $\{M_x\}$, the estimator variance $\Delta \phi^2$ is lower bounded by the inverse classical Fisher information (FI) associated with this strategy. The classical FI is defined by $F_{\phi} = \sum_x P(x \lvert\phi) \,\left[\partial_{\phi} \ln P(x \lvert\phi) \right ] ^2$ and captures the sensitivity of the probability distribution with respect to parameter changes. It is upper bounded by the quantum Fisher information (QFI), which contains a maximization over all measurement strategies. So overall we have the hierarchy $\Delta \phi^2 \geq 1/F_{\phi} \geq 1/\mathcal{F}_{\phi}$. Since the phase imprint and the spontaneous emission channel commute, one can just calculate the QFI for an unitary parameter imprint, that is \cite{Paris09}
\begin{equation}
    \label{eq:QFI_formula}
    \mathcal{F}_{\phi} = 4 \! \! \! \sum_{\substack{i < j \\ \lambda_i + \lambda_j > 0}} \frac{(\lambda_i - \lambda_j)^2}{\lambda_i + \lambda_j} \lvert \langle \chi_i \rvert S_z \lvert \chi_j \rangle \rvert^2,
\end{equation}
where $\{\lambda_i\}$ and $\{ \lvert \chi_i \rangle \}$ are the eigenvalues and eigenstates of $\rho_{\phi}(\eta)$. Alternatively, the QFI can be determined directly from its definition, i.e. $\mathcal{F_{\phi}} = \text{tr}( \rho_{\phi}(\eta) \, L_{\phi}^2 )$. Here, $L_{\phi}$ is the symmetric-logarithmic derivative (SLD) \cite{Paris09}, which is the quantum analogue of the classical score function. This operator can be obtained as an hermitean solution of 
\begin{equation}
    L_{\phi} \, \rho_{\phi}(\eta) + \rho_{\phi}(\eta) \, L_{\phi} = -2i [S_z, \rho_{\phi}(\eta)],
\end{equation}
and the lower bound of $\Delta \phi^2$ is always reachable by a measurement in the eigenbasis of the SLD \cite{Paris09, Kaufm25}. The QFI is additive for tensor product states \cite{Pezze18}, so for $\lvert \text{CSS} \rangle$ one has to evaluate the QFI of a time evolved single qubit and multiply by the size of the ensemble. It is straight forward to show that $\mathcal{F}_{\phi} = N e^{- \eta}$ in this case, and the lower bound $\Delta \phi^2 = 1/\mathcal{F}_{\phi}$ is exactly saturated by the standard Ramsey sequence, i.e. a collective measurement of $S_y$ and a linear phase estimator rescaled by the slope of the signal \cite{Kiel24}. For frequency estimation, the value $\mathcal{F}_{\omega} = \max_{\eta} \eta \, \mathcal{F}_{\phi}(\eta)$ is attained at the optimal interrogation time $T_{\text{opt}} = 1/\gamma$, such that $\eta_{\text{opt}} = \gamma \, T_{\text{opt}} = 1$. So the lower bound for frequency estimation is given by $\Delta \omega^2 = \mathcal{F}_{\omega}^{-1} = e/N$. This defines the uncertainty limit of classical strategies, and in the main section all bounds for frequency estimation are expressed as the quantum gain over this value.

We further benchmark SGSs to the uncertainty limits of the ``SSS protocol'', i.e. a standard Ramsey sequence with an SSS as the input state, and a measurement of $S_y$ combined with the standard linear estimator rescaled by the slope of the signal. Working in the Heisenberg picture, it is easy to show that the variance of such a standard Ramsey sequence is given by $\Delta \phi^2 = (N (e^{\eta} -1)/4 + \langle S_y^2 \rangle)/\langle S_x \rangle^2$ \cite{Kiel24}, where the expectation values are evaluated with respect to the initial state. For SSSs generated by OAT, the spin expectation values are known analytically, and we will always use SSSs which are rotated such that the squeezed axis is aligned along $y$ to get maximum benefit for the Hamiltonian $\propto S_z$ \cite{Kitag93}. For a given noise strength $\eta$, the phase uncertainty is a function of the squeezing parameter $\mu$ only, and we optimize $\Delta \phi^2(\mu)$ over this parameter. For frequency estimation, an additional optimization over the interrogation time is carried out to determine the lowest possible variance $\Delta \omega^2(\mu)$.

%% file: SUPP_MAT/3_PermInv.tex
\section*{M3. \enskip Permutational Invariant Dynamics}
\label{Meth:M3}
The total Hilbert space for a system of $N$ qubits is given by $\mathcal{H} \cong (\mathbb{C}^2)^{\otimes N}$. Using representation theory of the special unitary group, one can decompose the space $\mathcal{B}(\mathcal{H})$ of bounded operators on $\mathcal{H}$ into a direct sum of irreducible representations, where each invariant subspace is characterized by a fixed symmetry type \cite{Froewis14, Hart16}. The spontaneous emission master equation (\ref{eq:SE_MEQ}) and the unitary phase imprint describe tensor product dynamics, which is identical for all qubits in the ensemble. For an atomic clock, this means that all atoms see the same laser detuning during the Ramsey cycle, and all atoms have the same lifetime. Hence, this dynamics is permutational invariant, and different symmetry subspaces are not coupled during time evolution \cite{Hart16}. Particularly, a state with a fixed symmetry type stays within its symmetry sector. This motivates to restrict the dynamics to a certain symmetry subspace to make a numerical analysis feasible. We restrict to density operators from the permutational symmetric subspace $\mathcal{B}_{S_N}(\mathcal{H}) \subset \mathcal{B}(\mathcal{H})$, i.e. states $\rho$ which satisfy $P_{\sigma} \, \rho \, P_{\sigma}^{\dagger} = \rho$ for all permutations $\sigma \in S_N$, where $S_N$ is the symmetric group of $N$ objects. This restriction to completely symmetric states is physically motivated, but we are not aware of a proof that initial states which maximize the QFI have this symmetry. For some collective noise channels one can prove that metrologically optimal states are permutationally symmetric \cite{Dorner12}, but for local qubit channels there is in general no rigorous proof. The physical motivation is that the optimal initial states should have the same symmetry as the decoherence channel, so they should be permutationally invariant. Furthermore, a numerical maximization of the QFI within the full Hilbert space for qubit numbers $\mathcal{O}(1)$ shows that symmetric states perform metrologically much better than antisymmetric states or states of mixed symmetry type. Additionally, SSSs maximize the QFI in the asymptotic case $N \rightarrow \infty$ \cite{Kiel24}, which are also symmetric states of the ensemble.

The symmetric subspace $\mathcal{B}_{S_N}(\mathcal{H})$ is fully described by $\mathcal{O}(N^3)$ permutationally symmetric basis operators \cite{Froewis14, Hart16}. This polynomial scaling with the system size allows for an efficient numerical simulation of qubit master equations within that subspace, see e.g. the PIQS library in Python \cite{Shammah18}. Additionally, for the spontaneous emission master equation in Eq. (\ref{eq:SE_MEQ}), the dynamics is easily solvable algebraically in the permutationally symmetric operator basis \cite{Hart16}. Let $K_{m, m^{\prime}, d}$ be the permutationally symmetric basis operators with their respective quantum numbers $(m, m^{\prime}, d)$ as defined in reference \cite{Froewis14}. It is then straightforward to show that the spontaneous emission channel acts via \cite{Hart16}
\begin{align}
\label{eq:timeevol_Opbas}
	& \mathcal{E}_{\gamma T} \left( K_{m, m^{\prime}, d} \right) = e^{\mathcal{L}_{\text{SE}}T} \left( K_{m, m^{\prime}, d} \right) = e^{-N \frac{\gamma T}{2}} e^{-(m^{\prime} - m) \frac{\gamma T}{2}} \left( 1 - e^{- \gamma T} \right)^m \; \sum_{l = l_{\text{min}}}^{m} \; \beta_l \, K_{l, l + m^{\prime} - m, d}, \\[8pt] & \quad \text{with} \quad \beta_l = \left( e^{\gamma T} - 1 \right)^{-l} \, \binom{\frac{N + m - m^{\prime}}{2} - d - l}{m - l} \quad \text{and} \quad l_{\text{min}} = - \frac{N - m + m^{\prime}}{2} + d.  
\end{align}
Each initial state $\rho \in \mathcal{B}_{S_N}(\mathcal{H})$ is representable in the basis $\left \{ K_{m, m^{\prime}, d} \right \}$, and hence also the time evolved state. For the numerical maximization of the QFI it is further necessary to express permutationally symmetric states in a basis which allows to efficiently diagonalize the state, see Eq. (\ref{eq:QFI_formula}). We express $\rho \in \mathcal{B}_{S_N}(\mathcal{H})$ with respect to the Dicke basis $\{\lvert j,m,\alpha \rangle \}$ of $\mathcal{H}$, which is the eigenbasis of the collective spin operators $S_z$ and $S^2$. The coupled pseudospins take values $j = j_{\text{min}}, j_{\text{min}}+1, \dots, N/2$ with $j_{\text{min}} = 0 \left(\frac{1}{2}\right)$ for an even (odd) number of qubits, the respective projection quantum numbers satisfy $m = -j, -j+1, \dots , j$, and $\alpha = 1, 2, \dots, \Delta_j$ is the degeneracy index. In this basis, any $\mathcal{\rho} \in \mathcal{B}_{S_N}(\mathcal{H})$ decomposes into block diagonal form \cite{Chase08, Xu13, Froewis14}, where each block is characterized by a pseudospin $j$. Furthermore, blocks with different multiplicity indices $\alpha$ always have identical matrix elements, which considerably simplifies the numerical treatment. Overall, each symmetric density matrix decomposes as
\begin{equation}
    \label{eq:permSymm_BlockDiag}
    \rho = \bigoplus_{j = j_{\text{min}}}^{N/2} \left(\rho_j \right)^{\oplus \Delta_j} \quad \text{with multiplicity} \quad \Delta_j = \binom{N}{\frac{N}{2} - j} \;\frac{2j+1}{\frac{N}{2} + j + 1}.
\end{equation}
Consequently, any state $\rho \in \mathcal{B}_{S_N}(\mathcal{H})$ is representable by $\lfloor \frac{N}{2} + 1 \rfloor$ matrices $\rho_j$, where $\rho_j$ has the size $(2j + 1)^2$, which allows for efficient blockwise numerical diagonalization. For the numerical optimization of the QFI over the initial states, we analytically solve the spontaneous emission master equation in the symmetric operator basis, see Eq. (\ref{eq:timeevol_Opbas}) and reference \cite{Hart16}, and express the time evolved state $\rho_{\phi}(\gamma \, T)$ in block diagonal form as in Eq. (\ref{eq:permSymm_BlockDiag}). To carry out this particular basis transformation, we utilize the same technique as outlined in \cite{Froewis14}. The computational complexity to evaluate the QFI is then mainly determined by the time it takes to numerically diagonalize the individual matrices $\rho_j$.

%% file: SUPP_MAT/4_GlobalOpt.tex
\section*{M4. \enskip Numerical Maximization of the QFI}
\label{Meth:M4}
The optimal quantum interferometer (OQI) for phase estimation is defined as the global maximum of $\mathcal{F}_{\phi}$, see Eq. (\ref{eq:QFI_formula}), over permutationally symmetric initial states $\rho_0$. For frequency metrology, we analogously define the OQI as the maximum of $\mathcal{F}_{\omega} = \max_{\eta} \eta \, \mathcal{F}_{\phi}(\eta)$, with $\eta = \gamma \, T$, over all initial states. Hence, this contains an optimization over the duration $T$ of a single Ramsey interrogation. Since the QFI is a convex quantity \cite{Pezze18}, it suffices to optimize over pure initial states $\rho_0 = \lvert \psi_0 \rangle \! \langle \psi_0 \rvert$. Because of the block diagonal structure in Eq. (\ref{eq:permSymm_BlockDiag}), pure states within the subspace $\mathcal{B}_{S_N}(\mathcal{H})$ are given by superpositions of symmetric Dicke states, i.e. states with pseudospin $j = j_{\text{max}} = N/2$. We parametrize pure symmetric states by
\begin{equation}
    \label{eq:Global_Opt_InitState}
    \lvert \psi_0 \rangle = \sum_m c_m \; \lvert N/2, m \rangle,
\end{equation}
with $N+1$ complex coefficients $c_m \in \mathbb{C}$. Due to spontaneous emission, blocks with pseudospins $j < j_{\text{max}}$ get occupied during time evolution, but the final state $\rho_{\phi}(\gamma \, T)$ always has the block structure as in Eq. (\ref{eq:permSymm_BlockDiag}). So to calculate the QFI, one has to numerically diagonalize the time-evolved state blockwise and weight each contribution with the multiplicity of the corresponding symmetry subspace. Thus, the QFI is given by $\mathcal{F}_{\phi} = \sum_j \Delta_j \cdot \mathcal{F}_{\phi}^{(j)}$, where $\mathcal{F}_{\phi}^{(j)}$ is the QFI with respect to $\rho_j$, i.e. the non-normalized $j$-sector of the final state $\rho_{\phi}(\gamma \, T)$. Alternatively, one can calculate the QFI by directly solving the defining equation of the SLD blockwise for each $j$. This is a Sylvester equation which can be solved with standard numerical libraries. But note that this calculation via the SLD is numerically slower by roughly a factor of two than a computation of the QFI with the spectral formula in Eq. (\ref{eq:QFI_formula}).

The QFI is a function of the numerically obtained eigenvalues and eigenstates of the final state $\rho_{\phi}(\gamma \, T)$. Since the rank of this state heavily depends on the initial distribution $\{c_m \}$ and $T$, the eigenvalues will cross, degenerate, and hit zero during global optimization. Gradient-based algorithms are known to be prone to convergence problems under such situations and frequently stuck in suboptimal local extrema. For that reason, we utilize stochastic, gradient-free optimization algorithms to maximize the QFI. Particularly, we use differential evolution (DE), which is an evolutionary, population-based method for global optimization. After initialization of a number of population members (typically $400-600$ for the maximization of the QFI over $\{c_m \}$ for $N \gtrsim 50$ qubits), a mutation strategy suggests new trial population members. These trial members are typically generated by weighted differences of current population members. The population is then updated by those members which yield improved QFI values. This procedure is repeated until a stopping criterium given by the improvement of the QFI values between successive epochs is reached. We use the Python library DetPy \cite{Detpy26}, which offers 30 different variants of the DE algorithm. Various control parameters and optimizer settings can be tuned to explore different mutation strategies, as well as schemes to generate new trial population candidates. Within DetPy, the SHADE strategies were particularly useful to maximize the QFI and reach fast and consistent convergence. SHADE employs self-adaptive methods to learn useful mutation and selection strategies during optimization. In this way, the algorithm dynamically switches between local and global search during different stages of the optimization, and thus keeps the population diversity high to explore the full parameter space without premature convergence. Since DE algorithms are stochastic, it is necessary to carry out multiple runs for different initial populations to check consistency of the maximized QFI values, and to check if always optimal states with the same properties are returned. At least $15-20$ independent runs were carried out for each optimization result presented in the main sections.

As a further consistency check, we did some runs with the covariance matrix adaption (CMA-ES) strategy using the cma library in Python, which is one of the standard methods for noisy black box optimizations. This algorithm is also stochastic and population-based, but uses a very different strategy to generate population vectors compared to DE. At any epoch, population vectors are sampled according to a multi-dimensional Gaussian distribution and the QFI values are evaluated. Based on those results, the mean, width, and the covariance matrix of the Gaussian distribution are updated. In this way, the search cloud moves through the parameter space and eventually converges to a Gaussian centered roughly at the true global optimum.

In general, DE variants and also CMA-ES perform well and yield consistent results for the maximization of the QFI despite the large number of parameters for global optimization. Different stochastic algorithms converge consistently to the same global optimal states for the respective regimes of $N$, i.e. uGHZ-like states and spin GKP-like states, and we obtained values of $\mathcal{F}_{\phi}$ and $\mathcal{F}_{\omega}$ which are better than for other known state classes. Especially due to the large dimensionality of the problem, population-based strategies manifest their efficacy. For the global optimization over all symmetric states of a system with $N$ qubits, one has to maximize the QFI over the coefficients $\{ c_m \}$, i.e. $2(N+1)$ real parameters. Additionally, for frequency metrology with the cost function $\mathcal{F}_{\omega}$, one also has to determine the optimal Ramsey interrogation time $T$. The optimizer compares different state classes from the symmetric subspace, each with an individual optimal interrogation time. This results in many different local maxima, and a non-convex, complicated, and rugged optimization landscape. To check consistency, we further optimized just over the absolute values $\{ \lvert c_m \lvert \}$ where all relative phases are set to zero, which reduces the number of relevant parameters by a factor of $2$. The different optimization runs for $\{ \lvert c_m \rvert \}$ and $\{c_m\}$ yield approximately the same optimal QFI values in the corresponding regimes of $N$, the states are just trivially rotated by different angles about $z$ if relative phase are included. Of course, the optimization over $\lvert c_m \rvert$ converges faster. But we always found the same optimal state classes as in the full symmetric subspace, and we were not able to obtain noticeable better QFI values by including relative phases. Since the time-evolved states can be calculated analytically and are block diagonal, see Eq. (\ref{eq:permSymm_BlockDiag}), the computational complexity is mainly determined by the time it takes to diagonalize the individual matrices $\rho_j$. For e.g. $N = 125$, the final state is described by $63$ matrices $\rho_j$, where the largest matrix has pseudospin $j = 125/2$ and thus the matrix size $126 \times 126$. Diagonalizing all these matrices and calculating the QFI took $\lesssim 0.25$ seconds on a standard laptop. So a global optimization of the QFI was easily possible even for $N=125$ within a few hours, despite the large number of optimization parameters (more than $250$ for $N= 125$). With PIQS for example, this is not possible because it would take a few seconds to even determine the time evolved state numerically. Overall, we are confident to have found states performing close to the true global maximum of the QFI because of independent algorithmic agreement and convergence evidence. Different stochastic optimization strategies point to similar optimal QFI values and optimal states with similar properties, and they converge consistently for different random initial populations.

Finally, note that we used the same strategies also for the low-dimensional optimizations within the variational class in Eq. (\ref{eq:varclass}), and for the optimizations over the optimal gate parameters of the POVM for the phase readout scheme. For those optimizations, it was important to restrict the gate parameters to one period, e.g. $\mu_1 \in [0, \pi], \; \phi_x \in [-\pi, \pi], \; \text{and} \;\mu_2 \in [0, \pi]$ for the three gate parameters of the variational class, to simplify the optimization landscape and provide better and faster convergence. We additionally maximized the QFI for initial states created with many gate layers of OAT operations and rotations, since this can provide universality in the completely symmetric subspace \cite{Bond25}. But even for deep circuits with more than $15$ gates, no better solution than from the global optimization of the $\{ c_m \}$ could be obtained. We just found different versions of SGSs, but created with deeper circuits now and without any further improved metrological gain. We are confident that there are no materially better QFI values or state classes than the ones presented in the main section. Particularly, for $N$ large enough, we always observed optimal initial states with local grid structures on the Bloch sphere, either spin GKP states or SGSs.

%% file: SUPP_MAT/5_GKP_SpinGKP.tex
\section*{M5. \enskip GKP States and Spin GKP States}
\label{Meth:M5}

\textit{GKP States ---} Gottesman-Kitaev-Preskill (GKP) states are used for CV quantum error correction \cite{Gottesman01}, where logical qubits are encoded in the infinite dimensional Hilbert space of a harmonic oscillator. For the case of a single bosonic mode, that is $[a,a^{\dagger}] = \mathbb{I}$ with generalized quadratures $Q$ and $P$, ideal GKP states are defined as a superposition of quadrature eigenstates which are periodically displaced in a certain direction. Hence, ideal GKP states are an infinitely extended superposition of $\delta$-distributions in the respective quadrature eigenbases. These states show a comb structure in phase space with evenly spaced comb tips, such that a periodic lattice is formed. The Pauli algebra is then realized by displacement operators in the two quadrature directions, and similarly for the GKP stabilizer group used for syndrome extraction \cite{Gottesman01}. Therefore, a certain spacing of the comb tips is needed, e.g. $2 \sqrt{\pi}$ for the logical GKP states of the standard square GKP lattice. GKP states can be used for error correction via syndrome extraction in the usual way, and they are explicitly designed to correct for and protect against small displacement errors in the two generalized quadratures $Q$ and $P$. If the displacement error is smaller than half the spacing between adjacent comb peaks, then one can perfectly correct this error by shifting the state back to the nearest comb tip. Ideal GKP states are not normalizable and thus unphysical. To define physical GKP states, quadrature eigenstates have to be replaced by states with finite squeezing. Let $D(\alpha) = \exp(\alpha a^{\dagger} - \alpha^{*} a)$ be the bosonic displacement operator and $S(r) = \exp(r(a^2- a^{\dagger 2})/2)$ the bosonic squeezing operator, then GKP qubits ($\mu = 0,1$) can be defined as \cite{Gottesman01, Omanakuttan23}
\begin{equation}
    \label{eq:GKP}
    \left \lvert \mu \right \rangle \; \propto \; \sum_{j \in \mathbb{Z}} e^{-\frac{\pi \kappa^2}{2}(2j+\mu)^2} \; D\left((2j+\mu)\sqrt{\pi}/\sqrt{2}\right) \cdot S\left (-\ln(\Delta)\right) \left\lvert 0 \right \rangle,
\end{equation}
where $\lvert 0 \rangle$ is the bosonic vacuum state and $\Delta$ characterizes the squeezing strength. Here, the Gaussian envelope with width $\kappa$ acts as a cutoff to effectively superpose only a finite number of squeezed states. The Wigner function of a physical GKP qubit shows a highly structured grid in phase space, similar to a chessboard with clearly separated peaks, and this grid determines the error correction properties. Apart from correction of displacement errors, GKP states also have applications in displacement sensing \cite{Duiven17}, due to an effective squeezing in both quadratures locally around a comb peak. 
\newline
\newline
\newline

\textit{Spin GKP States ---} By analogy, GKP states of spin systems have been defined, which show grid structures on the Bloch sphere. This is done via the Holstein-Primakoff (HP) approximation in the limit of large spin $j$ \cite{Omanakuttan23}. Bosonic Fock states are identified with spin Dicke states, and the HP mapping is used to express spin observables by bosonic creation and annihilation operators. This constitutes a map from the unbounded bosonic phase space to the compact Bloch sphere of a spin. So for large spins, spin GKP states are created in a tangent plane of the Bloch sphere. For $j \rightarrow \infty$, this plane becomes the bosonic phase space such that spin GKP states converge to their bosonic GKP counterparts \cite{Omanakuttan23}. We look at the symmetric subspace of a system with $N$ qubits (i.e. $j = N/2$) and create spin GKP states with the same polarization as the coherent spin state $\lvert \text{CSS} \rangle$. So we identify bosonic excitations with spin Dicke states in the $x$-basis, and in particular the bosonic vacuum $\lvert 0 \rangle$ is identified with $\lvert N/2, N/2 \rangle_x = \lvert \text{CSS} \rangle$. For large $N$ and small excitations, the HP approximation yields $a^{\dagger} \approx S_-/\sqrt{N}$ and $a \approx S_+/\sqrt{N}$. Here, $S_-$ and $S_+$ refer to lowering or raising the collective spin projection $S_x$. Hence, spin GKP states are constructed by replacing bosonic operators in the definition (\ref{eq:GKP}) of a GKP state by spin operators \cite{Omanakuttan23}. Displacements turn into approximate collective rotations, e.g.
\begin{equation}
	\label{eq:rot_HP}
    D(\alpha) = e^{\alpha a^{\dagger} - \alpha^{*}a} \approx e^{-\frac{2i}{\sqrt{N}}(\text{Re}(\alpha)S_z - \text{Im}(\alpha)S_y)},
\end{equation}
and the bosonic squeezing approximately turns into a two-axis counter-twisting (TACT) interaction for spins,
\begin{equation}
	\label{eq:squeeze_HP}
    S(r) = e^{\frac{r}{2}\left(a^2 - a^{\dagger 2}\right)} \approx e^{\frac{r}{2} \frac{1}{N} \left(S_+^2 - S_-^{2}\right)}.
\end{equation}
Particularly, note that spin GKP states are defined such that the GKP stabilizer relations needed for syndrome detection are satisfied in the limit $N \rightarrow \infty$, which really makes spin GKP states the finite dimensional analogues for quantum computing with a collective spin system. This corresponds to a certain spacing of the grid lattice in order to realize the Pauli algebra via displacement operators in the asymptotic limit. Furthermore, for spin GKP states one can transport error correction algorithms, measurement strategies, and circuits from the bosonic system to the Bloch sphere via the HP mapping \cite{Omanakuttan23}. It is important to note that, since spin GKP states are defined in a tangent plane of the Bloch sphere, they are necessary polarized. This fundamentally distinguishes them from SGSs, see Fig. \ref{Fig:1} and Fig. \ref{Fig:3} in the main section, since the grid of an SGS extends once around the equator of the Bloch sphere due to the rotations in Eq. (\ref{eq:decomposeOATROT}) spanning angles up to $2 \pi$. Also, the stabilizer relations in general do not hold for SGSs, whereas spin GKP states are defined by a particular grid shape and spacing to approximately saturate the stabilizer relations. But similar as SGSs, spin GKP states are created by superposing certain versions of squeezed states placed at the equator of the Bloch sphere, which creates the fine, grid-like interference pattern. However, the Gaussian envelope in Eq. (\ref{eq:GKP}) imposes a cutoff such that the squeezed states are not placed around the full equator, but only on a spherical cap. This makes spin GKP states polarized on the Bloch sphere.
\newline
\newline
\newline

\begin{figure*}[t]
    \centering
    \includegraphics[width=0.9\columnwidth]{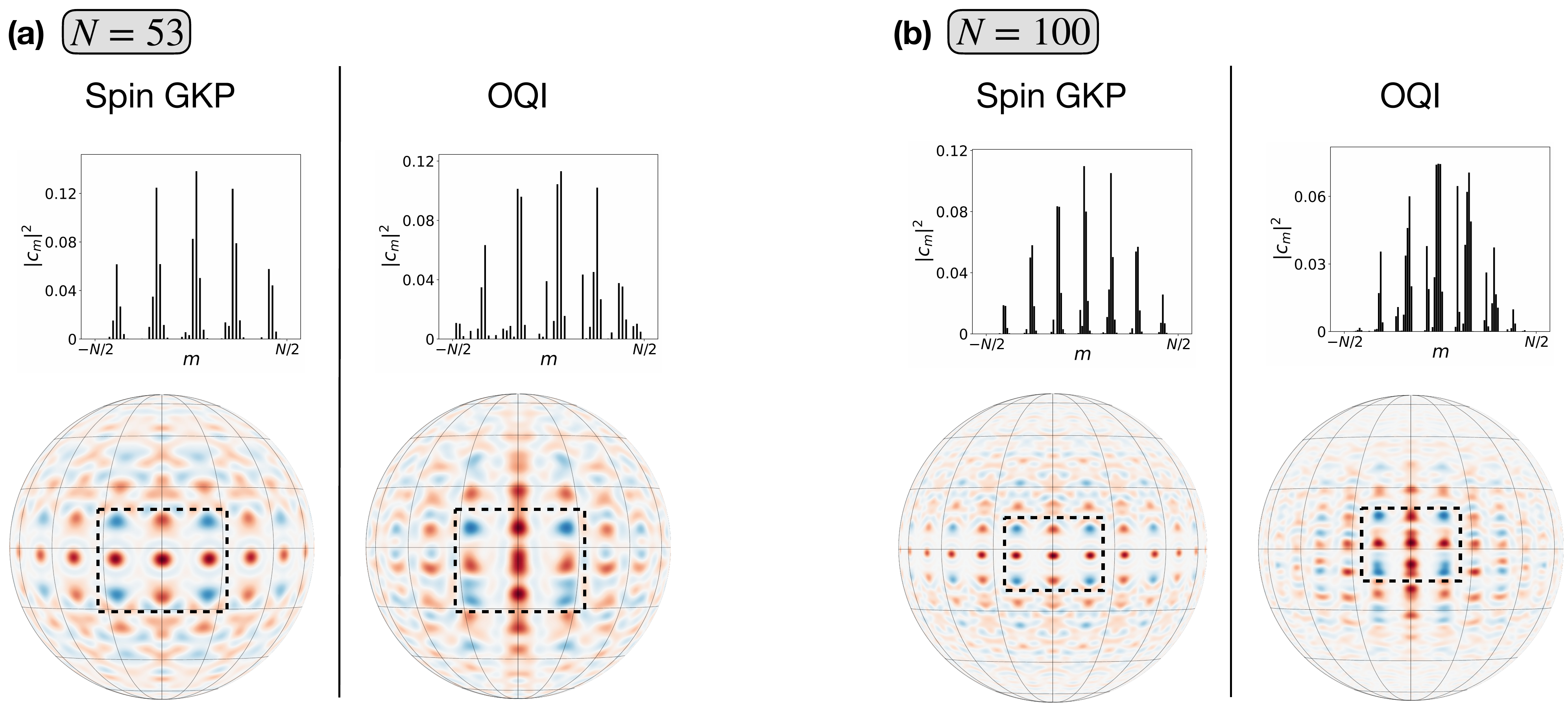}
    \caption{Initial state of the OQI (right column, here with a possible trivial rotation about $z$ to align the state along $x$) and optimal initial state within the spin GKP ansatz class (left column) for \textbf{(a)} $N=53$ qubits and \textbf{(b)} $N=100$ qubits. Optimal means that the QFI of frequency estimation, $\mathcal{F}_{\omega}$, is maximized and thus the estimation uncertainty is minimized. The same periodic grid structure exhibited by the OQI can be reproduced by spin GKP states, see the Wigner function plots (view plane: YZ) on the Bloch sphere and particularly the dashed boxed regions around the optimal working point at the equator, which have exactly the same size for the OQI and the spin GKP state. For larger $N$, the grid structures get finer and the states are more polarized and more tightly localized on a spherical cap of the Bloch sphere. Additionally, plots of the initial state populations $\lvert c_m \rvert^2$ are displayed. The optimal spin GKP state matches the OQI state remarkably closely in terms of the number of distinguished comb peaks, their positions along the $m$-axis, and the spacing between adjacent peaks.}
    \label{Fig:SUPP_GKP}
\end{figure*}

\textit{Global Optimality of Spin GKP States---} The OQI is defined as the boundary of the inaccessible region in Fig. \ref{Fig:3}, and the optimal initial states are obtained by a maximization of the QFI across all permutationally symmetric states $\lvert \psi_0 \rangle = \sum_m c_m \lvert N/2, m \rangle$. This sets the ultimate lower bound of frequency estimation. In the regime where SGSs are close to the OQI, i.e. $N \gtrsim 50$ for frequency metrology, the initial states of the OQI look very similar to spin GKP states, see Fig. \ref{Fig:SUPP_GKP}. And as discussed previously, this does not just hold for an optimization over $\lvert c_m \rvert$ without relative phases, but also for the full optimization over $c_m \in \mathbb{C}$, i.e. $2(N+1)$ optimization parameters. In the former case, the optimal initial states of the OQI are polarized along $x$, in the latter case the optimal states might be rotated about $z$ by some angle, which does not change the QFI value. The initial states of the OQI are polarized on the Bloch sphere and show a periodic grid pattern in phase space with clearly separated comb peaks. Plotting the initial state populations $\lvert c_m \rvert$ in the $z$-basis, one can see an oscillating comb pattern with overall envelope, and a comb period that depends on the number of qubits. This pattern roughly corresponds to the ``projection'' onto the $z$-axis and looks very similar to the marginal distributions of a bosonic GKP \cite{Gottesman01}, which are the projections onto the respective quadrature axes. To quantify how close the states of the OQI are to rigorously defined spin GKP states as in \cite{Omanakuttan23}, we maximize $\mathcal{F}_{\omega}$ for exact spin GKP initial states. For that purpose, we define the ``spin GKP ansatz class'' as polarized states of the form in Eq. (\ref{eq:GKP}), where all bosonic operators are replaced by spin operators via the HP approximation, see Eq. (\ref{eq:rot_HP}) and Eq. (\ref{eq:squeeze_HP}). So these states are parametrized by the squeezing strength $\Delta$, the width $\kappa$ of the overall envelope, and the GKP lattice configuration: For the definition of GKP states, totally general quadratures $Q$ and $P$ with $[Q, P] = i$ can be used to define the displacements along an arbitrary direction. These generalized quadratures are related to the standard quadratures $q = \left(a+a^{\dagger}\right)/\sqrt{2}$ and $p = -i\left(a-a^{\dagger}\right)/\sqrt{2}$ by a symplectic transformation. So a linear combination of $q$ and $p$, such that the canonical commutator is fulfilled, can be used to create grids on the Bloch sphere with different orientation and different grid spacings. This introduces three further optimization parameters, and additionally we optimize over the interrogation time. Overall, the spin GKP ansatz class can be used to approximate the initial states of the OQI, and this holds for different $N$, see Fig. \ref{Fig:SUPP_GKP}. To a very good approximation, the states of the OQI exhibit the same periodic structure in phase space as the optimal states within the spin GKP ansatz class. This is interesting because spin GKP states are defined by a certain grid spacing of the phase space lattice, such that the stabilizer relations and the Pauli algebra are satisfied in the large $N$ limit. For a given $N$, this is defined by the HP approximation, and spin GKP states show exactly the periodic grid structure dictated by this mapping. Consequently, the grid structure of spin GKP states has a very distinct physical meaning in terms of error correction and stabilizer operators. The OQI is obtained by maximizing the QFI over $c_m$, i.e. $2(N+1)$ parameters or $N+1$ parameters respectively if only the populations are considered, and the optimizer tunes these parameters to resemble the very same oscillating, periodic pattern of a spin GKP state.

%% file: SUPP_MAT/6_SGS_prot.tex
\section*{M6. \enskip Unravelling and SGS Protocol}
\label{Meth:M6}

\textit{Unravelling the Master Equation ---}
We write the full master equation as
\begin{equation}
    \dot{\rho} = -i \omega \left [ S_z, \rho \right ] + \frac{\gamma}{2} \sum_{i=1}^N 2 \sigma_-^{(i)} \rho \, \sigma_+^{(i)} - \rho \, \sigma_+^{(i)} \sigma_-^{(i)} - \sigma_+^{(i)} \sigma_-^{(i)} \rho = \mathcal{L}_{\text{eff}} \left( \rho \right) + \gamma \, \mathcal{J}\left[   \sigma_{-} \right] \left( \rho \right),
\end{equation}
where $\mathcal{L}_{\text{eff}}$ is the effective no-jump dynamics generated by the non-hermitian Hamiltonian $H_{\text{eff}}$,
\begin{equation}
    \mathcal{L}_{\text{eff}} \left( \rho \right) = -i \left(H_{\text{eff}} \, \rho - \rho \, H_{\text{eff}}^{\dagger}  \right) \quad \quad \text{with} \quad \quad H_{\text{eff}} = \left( \omega -i \gamma/2 \right)S_z - i \gamma N/4,
\end{equation}
and
\begin{equation}
    \mathcal{J}\left[ \sigma_{-} \right]\left( \rho \right) = \sum_m \sigma_-^{(m)} \rho \, \sigma_+^{(m)}
\end{equation}
describes a spontaneous-emission jump. Evolving $\rho_0$ in time, we have
\begin{equation}
    \rho(t) = e^{\mathcal{L}_{\text{eff}} t} \left ( \rho_0 \right )+ \int_{0}^t dt_1 \; e^{\mathcal{L}_{\text{eff}}(t-t_1)} \, \left ( \gamma \,\mathcal{J}\left[ \sigma_{-} \right]\left( \rho(t_1) \right) \right),
\end{equation}
which can be iterated in order to obtain a hierarchy of spontaneous-emission jump-trajectories between $t = 0$ and $t$, with a maximum number of $N$ decay events. Using commutation relations of $\mathcal{L}_{\text{eff}}$ and $\mathcal{J}$, the nested time integrals are easily solvable. The final state at time $T$ is then given by
\begin{equation}
    \rho_{\phi}(\gamma T) = \sum_{n=0}^N \, \rho_{\phi}^{(n)}(\gamma T) \quad \quad \text{with} \quad \quad \rho_{\phi}^{(n)}(\gamma T) = \frac{\left( 1 - e^{- \gamma T} \right)^n}{n!} e^{- i H_{\text{eff}} \, T} \mathcal{J}^n\left[ \sigma_{-} \right]\left( \rho_0 \right) \, e^{i H_{\text{eff}}^{\dagger}\,T} =: \mathcal{E}^{(n)}\left[\rho_0\right ]. 
    \label{eq:timeevol_state_METH}
\end{equation}
Hence, $\rho_{\phi}^{(n)}(\gamma T)$ is the unnormalized state conditioned on $n$ spontaneous-emission events. The probability that $n$ spontaneous-emission events occur is given by $p_n = \text{tr}\left(\rho_{\phi}^{(n)}(\gamma T)\right)$ such that $\sum_n p_n = 1$, and the phase $\phi = \omega T$ to be estimated is contained in $H_{\text{eff}} \, T$.
\newline
\newline
\newline
\newline
\newline
\textit{OAT Echo and Phase Kicks ---} The OAT operator $\mathcal{T}_{z}(\mu)$ satisfies
\begin{equation}
    \mathcal{T}_{z}(\mu)^{\dagger} \, \sigma_{-} \, \mathcal{T}_{z}(\mu) = e^{i \mu} e^{2 i \mu S_z} \, \sigma_{-},
\end{equation}
which can be easily shown in the tensor product basis. Consequently, the jump operator fulfills
\begin{align}
    \mathcal{J} \left[ \sigma_{-} \right]\left( \mathcal{T}_{z}(\mu) \, \rho \, \mathcal{T}_{z}(\mu)^{\dagger}  \right) &= \sum_{i = 1}^{N} \sigma_-^{(i)} \, \mathcal{T}_{z}(\mu) \, \rho \, \mathcal{T}_{z}(\mu)^{\dagger} \, \sigma_+^{(i)} = \mathcal{T}_{z}(\mu) \left (  \sum_{i=1}^{N} e^{2 i \mu S_z} \sigma_-^{(i)} \rho \, \sigma_+^{(i)} e^{- 2i \mu S_z}  \right ) \, \mathcal{T}_{z}(\mu)^{\dagger} \\[8pt]  
    &= \mathcal{T}_{z}(\mu) \left ( \mathcal{J} \left[ e^{2 i \mu S_z} \sigma_{-}  \right](\rho)   \right) \, \mathcal{T}_{z}(\mu)^{\dagger},
\end{align}
and similarly
\begin{align}
    \label{eq:jump_calculation_1}
    \mathcal{J}^n \left[  \sigma_{-} \right]\left( \mathcal{T}_{z}(\mu) \, \rho \, \mathcal{T}_{z}(\mu)^{\dagger}  \right) &= \mathcal{T}_{z}(\mu) \left ( \mathcal{J}^n \left[  e^{2 i \mu S_z} \sigma_{-}  \right](\rho)   \right) \, \mathcal{T}_{z}(\mu)^{\dagger} \\[8pt] & = \mathcal{T}_{z}(\mu) e^{2 i n \mu S_z} \left ( \mathcal{J}^n \left[   \sigma_{-} \right](\rho)   \right) e^{-2 in \mu S_z} \, \mathcal{T}_{z}(\mu)^{\dagger}
\end{align}
for $n$ successive applications. Hence, commuting the OAT past the operator $\mathcal{J}$ comes at the cost of a rotation kick, whose angle is determined by the strength of the OAT. Consequently, we receive $n$ rotation kicks if $n$ spontaneous emission events have happened. Note that this relation is particularly valid for the jump operator of spontaneous emission. E.g. for local dephasing noise generated by $\sigma_z$, the jump superoperator and the OAT commute because $\sigma_z$ and $S_z$ are both diagonal in the $z$-basis. 

Now, we take an SGS of the form $\lvert \psi_0 \rangle = \mathcal{T}_z(\mu_2) \lvert \xi \rangle$, where $\lvert \xi \rangle = \mathcal{R}_x(\phi_x) \mathcal{T}_z(\mu_1) \lvert \text{CSS} \rangle$ is an SSS, cf. Fig. \ref{Fig:1} (a). Applying an OAT echo to the time-evolved state in Eq. (\ref{eq:timeevol_state_METH}) yields
\begin{align}
    \mathcal{T}_{z}(\mu_2)^{\dagger} \,\rho_{\phi}(\gamma T) \, \mathcal{T}_{z}(\mu_2) &= \sum_{n=0}^N\, \mathcal{T}_{z}(\mu_2)^{\dagger} \,\rho_{\phi}^{(n)}(\gamma T) \, \mathcal{T}_{z}(\mu_2) = \sum_{n = 0}^N \, \bar{\rho}_{\phi}^{(n)}(\gamma T) \\[8pt]  &= \sum_{n=0}^N \frac{\left( 1 - e^{- \gamma T} \right)^n}{ \;n!} e^{- i H_{\text{eff}} \, T} \mathcal{T}_{z}(\mu_2)^{\dagger} \, \mathcal{J}^n \left[  \sigma_{-}\right] \left( \mathcal{T}_{z}(\mu_2) \lvert \xi \rangle \! \langle \xi \rvert \mathcal{T}_{z}(\mu_2)^{\dagger}\right) \, \mathcal{T}_{z}(\mu_2) \, e^{i H_{\text{eff}}^{\dagger}\,T} \\[8pt] &= \sum_{n=0}^N  \frac{\left( 1 - e^{-  \gamma T} \right)^n}{\,n!} e^{- i H_{\text{eff}} \, T} e^{2i n \mu_2 S_z} \, \mathcal{J}^n \left[  \sigma_{-} \right] \left( \lvert \xi \rangle \! \langle \xi \rvert \right) e^{-2i n \mu_2 S_z} \, e^{i H_{\text{eff}}^{\dagger}\,T} \\[8pt] &= \sum_{n=0}^N e^{2in\mu_2 S_z} \mathcal{E}^{(n)}(\lvert \xi \rangle \! \langle \xi \rvert) e^{-2in\mu_2 S_z}
\end{align}
Note that the term $\bar{\rho}_{\phi}^{(n)}(\gamma T)$ contains exactly $n$ rotation kicks. Since $\lvert \xi \rangle$ is an SSS, each jump term is like a decohered SSS, i.e. approximately Gaussian, and the overall state is a mixture of such states rotated to different positions on the Bloch sphere. The number of jumps is encoded by the position on the Bloch sphere, not just because the state effectively decays to the ground state at the south pole, but also because the OAT-echo induces rotation kicks along the latitude lines.
\newline
\newline
\newline
\newline
\newline
\textit{Ancillary-Assisted Estimation of the Phase ---} Generally, one can always saturate the QFI by a measurement in the eigenbasis of the SLD. But it is often not known how to express this measurement using a simple sequence of native, collective gates. In particular, because the noise acts locally on each qubit, the SLD is block-diagonal and blocks with different pseudospins $j$ get occupied, as discussed in Methods \hyperref[Meth:M3]{M3}. So to measure in the eigenbasis of the SLD, it is necessary to distinguish pseudospins $j \leq j_{\text{max}} = N/2$ and resolve different mixed symmetry types. This is completely unclear how to do in practice. Additionally, entanglement-enhanced collective measurements as defined in Eq. (\ref{eq:coll_obs}) cannot be used to approach the ultimate metrological bounds of SGSs either. Even entangling gates $\mathcal{U}_{\text{ent}}$ created with deep circuits of $\geq 20$ OAT gates and rotations, optimized over the gate parameters, do not suffice to approximate the QCRB and are far from optimal. 

But the unravelled expression suggests a strategy how to estimate the phase $\phi$ using the state $\mathcal{T}_{z}(\mu_2)^{\dagger} \,\rho_{\phi}(\gamma T) \, \mathcal{T}_{z}(\mu_2)$, where the OAT echo is already included. The jump contributions are roughly given by Gaussian states, but each of those states is rotated by a different angle about the $z$-axis which depends on the number of spontaneous-emission events. So the idea is to use a sequential measurement, which determines the number $n$ of jumps, and afterwards uses the $n$-conditioned state for the phase readout. For this purpose, we introduce an ancillary system, which effectively acts as a tag for the number of decay events via
\begin{equation}
 \sum_{n} \mathcal{T}_{z}(\mu_2)^{\dagger} \,\rho_{\phi}^{(n)}(\gamma T) \, \mathcal{T}_{z}(\mu_2) \; \otimes \; \lvert n \rangle \! \langle n \rvert,
\end{equation}
such that the number of jumps can be determined by a measurement on this extra system, without disturbing the phase information which is encoded on the clock atoms. For the ancillary system, we take a spin $j$ with spin operators $S_{z,A}$ and $S_{A}^2$, which is e.g. realized by the permutationally symmetric Dicke states of $N_A = 2j$ identical ancillary qubits. Let $\rho_{\text{in}} = \lvert \xi \rangle \! \langle \xi \rvert \otimes \rho_A$ with an SSS $\lvert \xi \rangle$ as defined above be the initial state of the combined system, then we use
\begin{equation}
    \mathcal{T}_{z,\, \text{tot}}(\mu_2) = e^{i \mu_2 \left( S_z + \epsilon S_{z, A} \right)^2} = e^{i \mu_2 S_z^2} e^{i \mu_2 \epsilon^2 S_{z, A}^2} e^{2 i \mu_2 \epsilon S_z \otimes S_{z,A}}
\end{equation}
to entangle both subsystems. For $\epsilon = 1$, this would just be a collective OAT acting on the combined system. Now, the phase imprint and the spontaneous emission channel act only on the clock, i.e.
\begin{equation}
    \mathcal{E}_{\text{tot}}\left(\mathcal{T}_{z,\, \text{tot}}(\mu_2) \, \rho_{\text{in}} \, \mathcal{T}_{z,\, \text{tot}}(\mu_2)^{\dagger} \right) =   \left ( \mathcal{E}_{\scriptscriptstyle \gamma T} \otimes \mathbb{I} \right)  \left ( \mathcal{E}_{\phi} \otimes \mathbb{I} \right)\left(\mathcal{T}_{z,\, \text{tot}}(\mu_2) \, \rho_{\text{in}} \, \mathcal{T}_{z,\, \text{tot}}(\mu_2)^{\dagger}\right).
\end{equation}
Although the noise channel acts trivially on the ancillary system, due to entanglement with the clock the information about the number of decay events is encoded on this extra system too. Similarly as in Eq. (\ref{eq:jump_calculation_1}), we have
\begin{align}
    \mathcal{J}^n \left[ \sigma_{-} \right]\left( e^{2i \mu_2 \epsilon S_z \otimes S_{z,A}} \, \rho_{\text{in}} \, e^{-2i \mu_2 \epsilon S_z \otimes S_{z,A}}  \right) =  e^{2i \mu_2 \epsilon S_z \otimes S_{z,A}} \, e^{2 i n \mu_2 \epsilon S_{z,A}}  \left ( \mathcal{J}^n \left[  \sigma_{-} \right](\rho_{\text{in}})   \right) \, e^{-2 i n \mu_2 \epsilon S_{z,A}} \, e^{-2i \mu_2 \epsilon S_z \otimes S_{z,A}}.
\end{align}
Consequently, a $\mathcal{T}_{z,\, \text{tot}}(\mu_2)$-echo yields
\begin{align}
    &\mathcal{T}_{z,\, \text{tot}}(\mu_2)^{\dagger} \; \mathcal{J}^n \left[ \sigma_{-} \right]\left( \mathcal{T}_{z,\, \text{tot}}(\mu_2) \, \rho_{\text{in}} \, \mathcal{T}_{z,\, \text{tot}}(\mu_2)^{\dagger}  \right) \, \mathcal{T}_{z,\, \text{tot}}(\mu_2) \\[8pt] & \quad \quad \quad = e^{2i n \mu_2   S_z} \, e^{2 i n \mu_2 \epsilon S_{z,A}}  \left ( \mathcal{J}^n \left[ \sigma_{-} \right](\rho_{\text{in}})   \right) \, e^{-2 i n \mu_2 \epsilon S_{z,A}} \, e^{-2in \mu_2 S_z },
\end{align}
and the transformed state is given by
\begin{align}
    \label{eq:echo_state}
    \rho_{\phi, \text{echo}}(\gamma T) &=\mathcal{T}_{z,\, \text{tot}}(\mu_2)^{\dagger} \; \mathcal{E}_{\text{tot}}\left(\mathcal{T}_{z,\, \text{tot}}(\mu_2) \, \rho_{\text{in}} \, \mathcal{T}_{z,\, \text{tot}}(\mu_2)^{\dagger} \right) \, \mathcal{T}_{z,\, \text{tot}}(\mu_2) \\[5pt] & = \sum_{n=0}^N \bar{\rho}_{\phi}^{(n)}(\gamma T) \otimes \; e^{2in \mu_2 \epsilon S_{z,A}} \, \rho_A \,e^{-2in \mu_2 \epsilon S_{z,A}},
\end{align}
where $\bar{\rho}_{\phi}^{(n)}(\gamma T)$ are exactly the same conditioned states as defined in Eq. (\ref{eq:njumpcontrib_NOECHO}). This is a separable state, and the ancillary system contains information about the number of jumps. Now, we prepare $\rho_A$ in a superposition of $n_0 < N$ Dicke eigenstates of $S_{z, A}$, i.e. $\rho_A = \lvert \psi_A \rangle \! \langle \psi_A \rvert$ with $\lvert \psi_A \rangle = n_0^{-1/2} \sum_{n = 0}^{n_0 - 1} \; \lvert n \rangle $. This yields
\begin{equation}
    e^{2in \mu_2 \epsilon S_{z,A}} \, \lvert \psi_A \rangle = \frac{1}{\sqrt{n_0}} \sum_{k = 0}^{n_0 - 1} e^{2 i n k \mu_2 \epsilon} \; \lvert k \rangle
\end{equation}
up to a trivial global phase. For the choice $\mu_2 \epsilon = \pi/n_0$, this is exactly a quantum Fourier transform (QFT) \cite{NielsenChuang} with respect to the Dicke basis states $\{ \lvert 0 \rangle, \lvert 1 \rangle, \dots, \lvert n_0 \rangle \}$. Hence, applying an inverse QFT to the ancillary system for $\mu_2 \epsilon = \pi/n_0$ and a large enough $n_0$ yields
\begin{equation}
    \label{eq:final_state_approx}
    \rho_{\phi,\text{final}}(\gamma T) =\left ( \mathbb{I} \otimes \mathcal{U}_{\text{QFT}}^{-1} \right )  \, \rho_{\phi, \text{echo}}(\gamma T) \, \left( \mathbb{I} \otimes \mathcal{U}_{\text{QFT}}^{} \right) \approx \sum_{n=0}^{n_0-1} \bar{\rho}_{\phi}^{(n)}(\gamma T) \, \otimes \, \lvert n \rangle \! \langle n \rvert.
\end{equation}
Note that the QFT acts modulo $n_0$, so the exact expression is
\begin{equation}
    \rho_{\phi,\text{final}}(\gamma T) = \sum_{n=0}^{n_0-1} \, \left( \bar{\rho}_{\phi}^{(n)}(\gamma T) \; + \; \bar{\rho}_{\phi}^{(n_0 + n)}(\gamma T) \; + \; \bar{\rho}_{\phi}^{(2 n_0 + n)}(\gamma T) \; + \; \dots\right)\; \otimes \; \lvert n \rangle \! \langle n \rvert.
\end{equation}
So the idea is to use $n_0$ as a cutoff and just work with the jump terms that contribute significantly to arrive at Eq. (\ref{eq:final_state_approx}) to a good approximation. Of course, we choose $n_0$ as small as possible to include as few additional resources as possible. In our case, the probability $p_k = \text{tr}\left(\bar{\rho}_{\phi}^{(k)}(\gamma T)\right)$ of $k$ jumps is small for large $k$. For an SGS of $55$ qubits evolved to its optimal interrogation time, we e.g. have $p_k < 10^{-4}$ for $k>20$, which defines a suitable cutoff $n_0$. The final state is then measured in the Dicke basis of the ancillary system. For a result $n$, the conditioned unnormalized state is $\bar{\rho}_{\phi}^{(n)}(\gamma T) \, \otimes \, \lvert n \rangle \! \langle n \rvert$, so a single jump term is isolated and we use a suitable measurement on the clock system to estimate the phase. 

For the measurement on the clock atoms, we use the observable
\begin{equation}
    X^{(n)} = \mathcal{U}_{\text{ent}}^{(n)} \, S_z \; \mathcal{U}_{\text{ent}}^{(n) \dagger} = \sum_M M \cdot \mathcal{U}_{\text{ent}}^{(n)} \, \mathbb{P}_M \; \mathcal{U}_{\text{ent}}^{(n) \dagger} =: \sum_M M \cdot X_M^{(n)}
\end{equation}
with outcomes $M \in \{ -N/2, -N/2 + 1, \dots, N/2 \}$ and projectors $\mathbb{P}_M$. This amounts to applying an entangling unitary, which depends on the number $n$ of spontaneous-emission events, before measuring the population difference on the clock system. Each entangling unitary $\mathcal{U}_{\text{ent}}^{(n)}$ is created with a low-depth variational circuit, which consists of collective rotations about arbitrary axes and OAT,
\begin{equation}
    \mathcal{U}_{\text{ent}}^{(n)} = \mathcal{R}_{\vec{k}}(\theta) \; \mathcal{T}_{z}(\mu) \; \mathcal{R}_{\vec{l}}(\chi).
    \label{eq:gate_layers_1}
\end{equation}
The metrological performance is quantified by the classical Fisher information (FI). Let $p_{n,M}(\phi) = \text{tr}\left( X_M^{(n)} \;\bar{\rho}_{\phi}^{(n)}\right)$ be the probability that $n$ jumps occurred and afterwards $M$ is measured, then the classical FI is
\begin{align}
    F_{\phi} &= \sum_{n,M} \frac{\left( \partial_{\phi} \, p_{n,M}(\phi) \right)^2}{p_{n,M}(\phi)} = \sum_n \sum_M \frac{\left( \partial_{\phi} \; \text{tr} \left( X_M^{(n)} \;  \bar{\rho}_{\phi}^{(n)} \right) \right)^2}{ \text{tr} \left( X_M^{(n)} \;  \bar{\rho}_{\phi}^{(n)} \right)} = \sum_n \sum_M \frac{ \text{Re}\left( \text{tr} \left( X_M^{(n)} \;  \bar{\rho}_{\phi}^{(n)} \; L_{\phi}^{(n)} \right) \right)^2}{ \text{tr} \left( X_M^{(n)} \;  \bar{\rho}_{\phi}^{(n)} \right)},
\end{align}
where the SLDs are defined as hermitian solutions of
\begin{equation}
    L_{\phi}^{(n)} \, \bar{\rho}_{\phi}^{(n)} \, + \, \bar{\rho}_{\phi}^{(n)} \, L_{\phi}^{(n)} = 2 \, \partial_{\phi} \, \bar{\rho}_{\phi}^{(n)} = -2 i \left[ S_z, \; \bar{\rho}_{\phi}^{(n)} \right].
\end{equation}
Upon optimizing the classical FI over the gate parameters, we find that a low-depth circuit is already enough to come close to the QCRB of SGSs and yield a gain compared to SSSs, see Fig. \ref{Fig:4}. The optimizations show that the OAT strength $\mu$ is small ($\ll 1$) and does not vary much with $n$, but the rotation parameters and axes are very different for different jump terms. This is because the rotation parameters have to be chosen such that the right position of the respective jump term on the Bloch sphere is addressed. To close the tiny remaining gap to the QCRB of SGSs, more gate layers e.g. of the form in Eq. (\ref{eq:gate_layers_1}) can be used.

%% file: SUPP_MAT/7_StateDist.tex
\section*{M7. \enskip Distinguishing Quantum States}
\label{Meth:M7}
In this section, we analyze the unravelled, time-evolved SGS in Eq. (\ref{eq:unravel_basic}) within the framework of quantum state discrimination to quantify the distinguishability of the jump terms $\rho^{(n)}_{\phi}(\gamma T)$. The general setting is the following: Let $\{ \rho_n \}$ be a set of normalized quantum states, each occurring with a certain probability $p_n $, such that the total ensemble state is $\rho = \sum_n p_n \, \rho_n$. Given a state $\rho_n$ chosen from the ensemble, we want to determine which state we have. So the goal is to correctly identify the index $n$ with high probability using some POVM $\{ \Pi_k \}$. The probability of being given the state $\rho_n$, but measure index $k$, is $P_n(k) = \text{tr}(\Pi_k \, \rho_n)$. Consequently, the success probability of state discrimination using the measurement $\{ \Pi_k \}$ is defined by
\begin{equation}
	\label{eq:Psucc}
    P_{\text{succ}} = \sum_n p_n \, P_n(n) = \sum_n p_n \, \text{tr} \left( \Pi_n \, \rho_n \right).
\end{equation}
A measurement strategy with $P_{\text{succ}} = 1$ can be used to discriminate all states with certainty. Finding a measurement which maximizes $P_{\text{succ}}$ for a given set of states and probabilities is a convex optimization problem. One has to optimize over POVM elements, i.e. with constraint $\sum_k \, \Pi_k = \mathbb{I}$ and $\Pi_k \geq 0$ for all $k$. This is a semidefinite program (SDP), so it is guaranteed to converge to a global optimum and numerically efficiently solvable (e.g. with CVXPY \cite{CVXPY16} in Python). Heuristically, the ``pretty good measurement'' (PGM) \cite{Hausladen94} is a strategy which performs remarkably close to the globally optimal measurement in many cases. The POVM elements of the PGM are defined by
\begin{equation}
	\label{eq:PGM}
    \Pi_k = p_k \; \rho^{-1/2} \; \rho_k \; \rho^{-1/2}.
\end{equation}
Here, the inverse is defined only on the support of $\rho$. Pairwise orthogonal states, that is $\text{tr} \left( \rho_n \, \rho_m \right) = 0$ for $n \neq m$, are perfectly distinguishable and the success probability  of the PGM is $P_{\text{succ, PGM}} = 1$ in this case \cite{Hausladen94}. So an optimal POVM for mutually orthogonal states is given by the PGM. In general, we have the hierarchy $P_{\text{succ, PGM}} \leq P_{\text{succ, max}} \leq 1$, and equality holds for perfectly orthogonal states $\{ \rho_n \}$. Consequently, the success probability of state discrimination also quantifies how pairwise orthogonal the states $\rho_n$ are. 

\begin{figure*}[t]
    \centering
    \includegraphics[width=0.7\columnwidth]{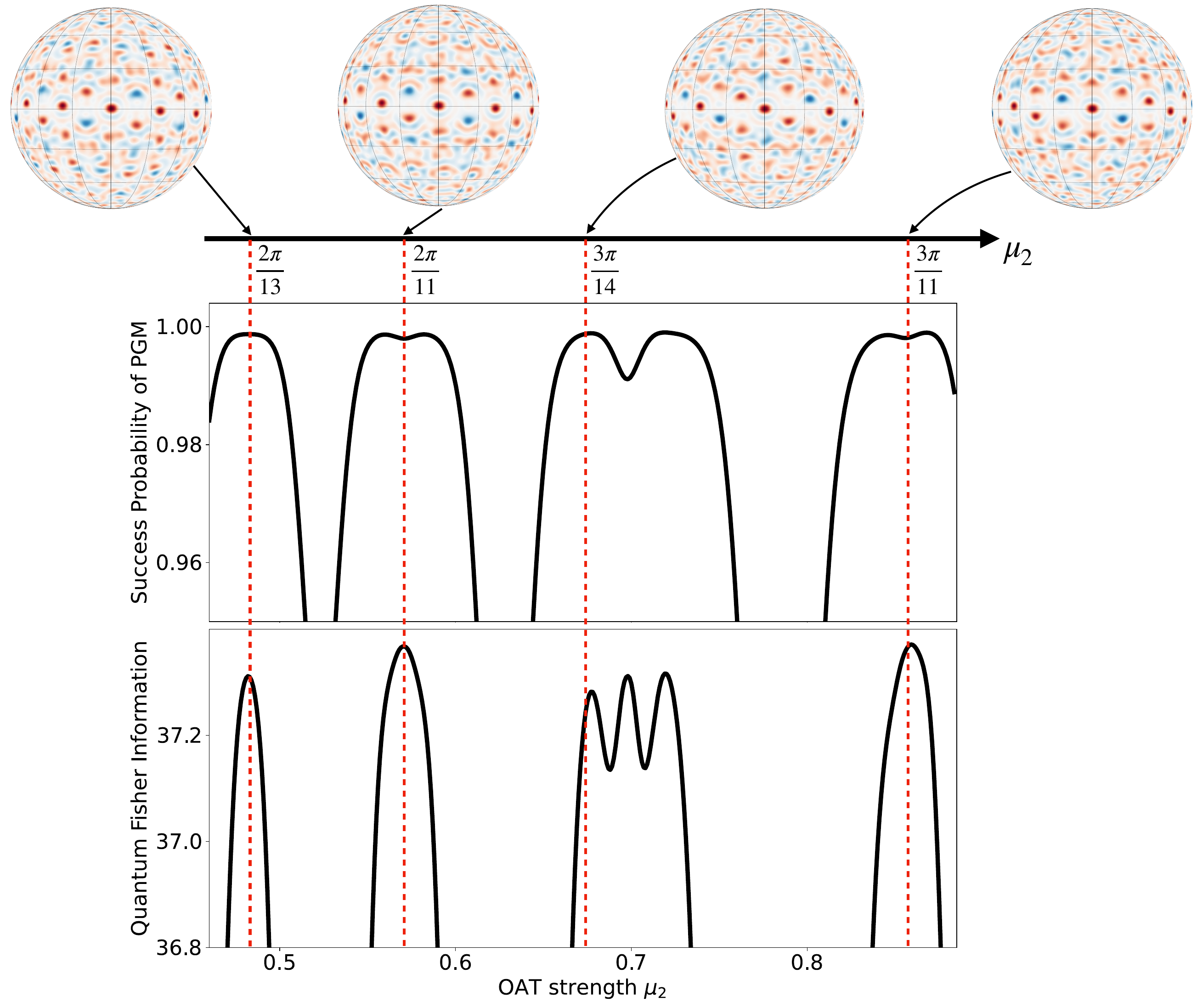}
    \caption{Success probabilities of the PGM and QFI values of phase estimation for different initial states of the form $\lvert \psi_0 \rangle = \mathcal{T}_{z}(\mu_2) \lvert \xi \rangle$, where $\lvert \xi \rangle$ is an SSS. Here, the number of qubits is $N=55$ and all states are evolved for the noise strength $\eta = \gamma T = 0.25$. For all $\mu_2$, the initial SSS is $\lvert \xi \rangle = \mathcal{R}_x(\phi_x)\mathcal{T}_{z}(\mu_1) \lvert \text{CSS} \rangle$ with fixed parameters $(\mu_1, \phi_x) = (0.073, -1.310)$. The success probability is evaluated with respect to the POVM of the PGM in Eq. (\ref{eq:PGM}). One can see that at local maxima of the QFI, also $P_{\text{succ, PGM}}$ is close to its maximum value. Furthermore, the Wigner functions corresponding to some exemplary local maxima are displayed on the Bloch sphere (view plane: YZ). In all cases, the grid is clearly visible.}
    \label{Fig:SUPP_DIST}
\end{figure*}

For time-evolved SGSs, the unnormalized states conditioned on $n$ spontaneous-emission events are $\rho^{(n)}_{\phi}(\gamma T)$, see Eq. (\ref{eq:unravel_basic}). The goal is to distinguish the number of decay events that have happened, i.e. measure the jump index $n$. Evaluating the success probability with respect to the PGM measurement, we find that SGSs have $P_{\text{succ, PGM}} \approx 1$ and hence allow for near perfect distinguishability of the jump terms, see Fig. \ref{Fig:SUPP_DIST}. The success probabilities in the figure are shown for $N=55$ and initial states of the form $\lvert \psi_0 \rangle = \mathcal{T}_{z}(\mu_2) \lvert \xi \rangle$ with different $\mu_2$, where $\lvert \xi \rangle$ is a fixed SSS. Furthermore, all the states are evolved for $\eta = \gamma T = 0.25$. Due to this large noise strength, there is a significant number of jump terms which contribute to the dynamics. For $N=55$ in the figure, we e.g. have $p_n = \text{tr} \left ( \rho^{(n)}_{\phi}(\gamma T) \right) < 10^{-4}$ for $n > 20$. And all those relevant, non-negligible  jump contributions are almost perfectly distinguishable by a measurement. One can see that $P_{\text{succ, PGM}} \approx 1$ for an SGS, irrespective of the particular global shape of the grid structure. This is of course generally not true for other initial state classes like CSSs, SSSs, or for random permutationally symmetric states. For example, SSSs have jump contributions with significant overlap such that typically $P_{\text{succ, PGM}} \approx 0.3$. Due to $P_{\text{succ, PGM}} \approx 1$, the jump contributions of SGSs are almost perfectly distinguishable and mutually orthogonal. This orthogonality can also be seen from the time-evolved state in Eq. (\ref{eq:njumpcontrib_NOECHO}), where the OAT echo has already been applied. The Hilbert-Schmidt inner product $\text{tr}(A \, B^{\dagger})$ and the success probability $P_{\text{succ}}$ are both invariant under unitary transformations of all states, so one can trivially apply such an echo. After the OAT echo, the jump contributions of SGSs are given by decohered SSSs, polarized along different directions on the Bloch sphere. Plotting the Wigner functions of the individual jump terms, one can see that these states are roughly given by different ellipses in phase space, which have almost no overlapping support on the Bloch sphere, so they are approximately pairwise orthogonal. 

Additionally, the QFI values for phase estimation are displayed in Fig. \ref{Fig:SUPP_DIST} for reference. This clearly shows the connection between the grid structure of SGSs, their large QFI values, and the distinguishability of the jump terms. Different versions of SGSs created with suitable $\mu_2 = p \pi/l$ are located at local maxima of the QFI, and they also yield success probabilities close to the maximum $P_{\text{succ, PGM}} \approx 1$. So the terms $\rho^{(n)}_{\phi}(\gamma T)$ are already pairwise orthogonal, and it is not necessary to ``artificially'' construct orthogonal jump terms as done in Eq. (\ref{eq:final_ancillaTag}) by using ancillary degrees of freedom. This suggests that there should be strategies using only the clock system, which first determine the number of decay events and then read out phase information on these conditioned states. Unfortunately, the canonical Lüders measurement $\sqrt{\Pi_k}$ generated by the POVM elements of the PGM cannot be used, since this does not commute with the phase imprint and thus destroys the phase sensitivity. Designing a sequential measurement that only uses the clock system requires more refined strategies.

Note that the orthogonality of the jump terms $\rho^{(n)}_{\phi}(\gamma T)$ can also be characterized by the ultimate convex bound \cite{Pezze18} of the QFI. Generally, the QFI of a state $\rho = \sum_m p_m \, \rho_m$ is upper bounded by $\sum_m p_m \, \mathcal{F}^{(m)}$, where $\mathcal{F}^{(m)}$ is the QFI of the state $\rho_m$. If all the $\rho_m$ are pairwise orthogonal, then exactly this upper bound is reached. Since the QFI is unitary invariant, one can directly see from Eq. (\ref{eq:njumpcontrib_NOECHO}) that this upper bound is completely independent of $\mu_2$. So for any $\mu_2$ in Fig. \ref{Fig:SUPP_DIST} we have the same convex bound. But by choosing a suitable $\mu_2$, one can make all the jump terms pairwise nearly orthogonal to actually come close to the upper convex bound. And the corresponding states are SGSs with a grid structure in phase space.